\documentclass[twocolumn,showpacs,preprintnumbers,amsmath,amssymb,floatfix]{revtex4}

\usepackage{graphicx}
\usepackage{dcolumn}
\usepackage{bm}

\newenvironment{wmatrix}{\left(\begin{array}[c]{cc}}{\end{array}\right)}
\newenvironment{wlmatrix}{\begin{array}[c]{cc}}{\end{array}}
\newcommand{\nmat}[4]{\begin{wlmatrix} #1 & #2 \\ #3 & #4 \end{wlmatrix}}

\begin{document}

\title{DC-transport in superconducting point contacts: a full counting 
statistics view}

\author{J.C.~Cuevas}
\affiliation{%
Institut f\"ur Theoretische Festk\"orperphysik,
Universit\"at Karlsruhe, D-76128 Karlsruhe, Germany
}%

\author{W.~Belzig}
\affiliation{
Department of Physics and Astronomy, University of Basel,
Klingelbergstr.82, CH-4056 Basel, Switzerland 
}%

\date{\today}

\begin{abstract}
  We present a comprehensive theoretical analysis of the dc transport
  properties of superconducting point contacts. We determine the full
  counting statistics for these junctions, which allows us to calculate
  not only the current or the noise, but all the cumulants of the
  current distribution. We show how the knowledge of the statistics of
  charge transfer provides an unprecedented level of understanding of
  the different transport properties for a great variety of situations.
  We illustrate our results with the analysis of junctions between BCS
  superconductors, contacts between superconductors with pair-breaking
  mechanisms and short diffusive bridges.  We also discuss the
  temperature dependence of the different cumulants and show the
  differences with normal contacts.
\end{abstract}


\maketitle

\section{Introduction}

The current-voltage (I-V) characteristics of superconducting contacts
have been the subject of investigation during the last four decades. 
The first experimental analyses were performed in tunnel
junctions~\cite{Giaever1962}. In this case the current inside the
superconducting gap is suppressed, and the results can be accurately
described with the BCS theory~\cite{BCS}. However, very often a
significant current is observed in the subgap region, which cannot be
explained with the simple tunnel theory. The first anomalies were
reported by Taylor and Burstein~\cite{Taylor1963} who noticed a small
onset in the current when the applied voltage $V$ was equal to the
energy gap, $\Delta/e$, in a tunneling experiment between two equal
superconductors. Relatively soon afterwards it was
apparent~\cite{Yanson1965,Marcus1966} that not only is there an anomaly
in the current at $eV=\Delta$, but in fact at all submultiples
$2\Delta/n$, where $n$ is an integer. This set of anomalies is referred
to as \emph{subharmonic gap structure} (SGS), and its temperature and
magnetic field dependence were thoroughly
characterized~\cite{Rowell1968,Bright1969,Giaever1970}.

\begin{figure}
\includegraphics[width=\columnwidth,clip=]{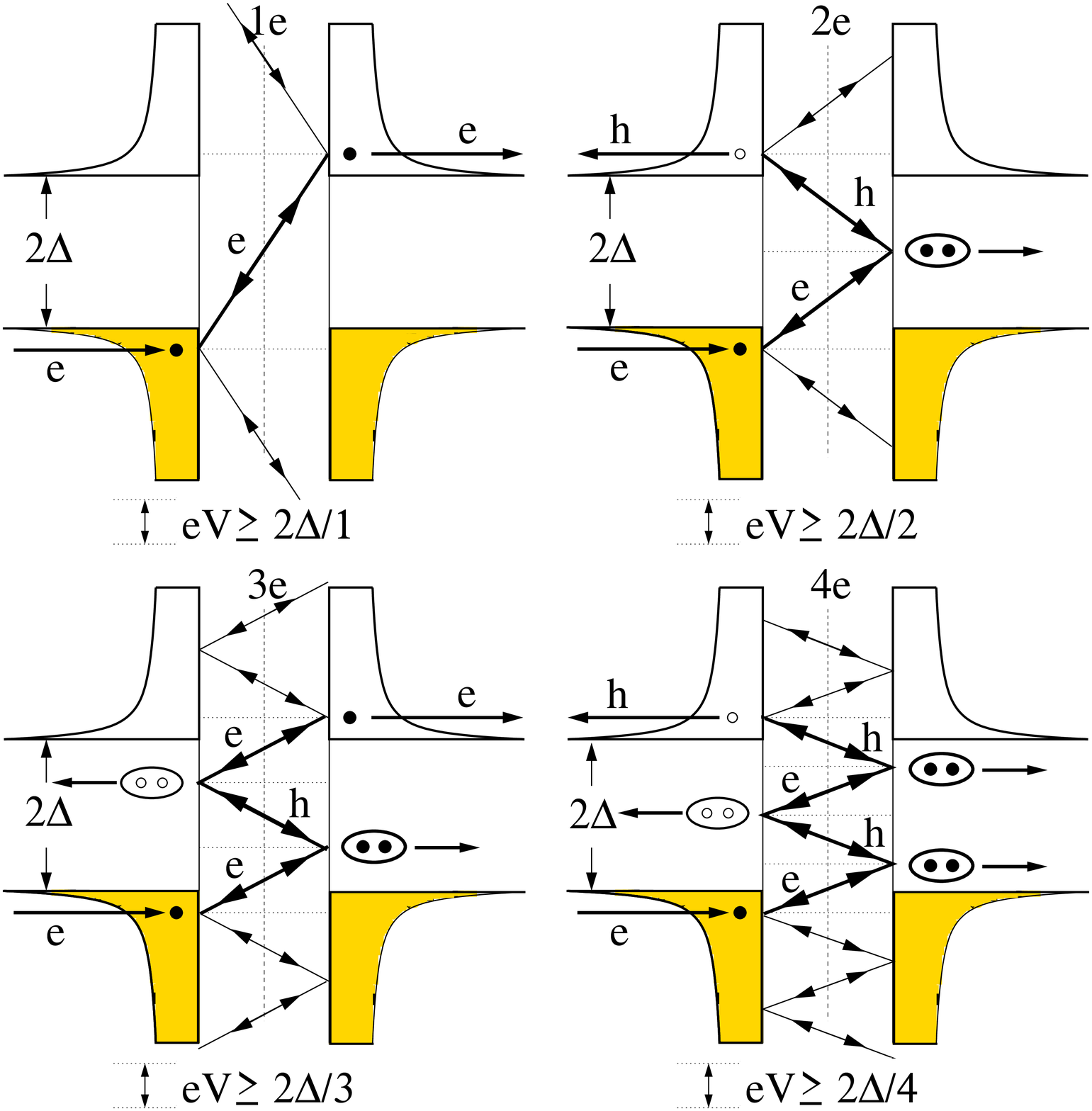}
\caption{\label{MAR} Schematic representation of the MARs for BCS
  superconductors with gap $\Delta$. We have sketched the density of
  states of both electrodes. In the upper left panel we describe the
  process in which a single electron tunnels through the system
  overcoming the gap due to a voltage $eV \ge 2\Delta$. The other panels
  show MARs of order $n=2,3,4$. In these processes an incoming electron
  at energy $E$ undergoes at least $n-1$ Andreev reflections to finally
  reach an empty state at energy $E+neV$. In these MARs a charge $ne$ is
  transferred with a probability, which for low transparencies goes as
  $T^n$. At zero temperature they have a threshold voltage
  $eV=2\Delta/n$. The arrows pointing to the left in the energy
  trajectories indicate that a quasiparticle can be normal reflected.
  The lines at energies below $E$ and above $E+neV$ indicate that after
  a detour a quasiparticle can be backscattered to finally contribute to
  the MAR of order $n$.}
\end{figure}

The first theoretical attempt to explain the SGS was done by Schrieffer
and Wilkins~\cite{MPT}, who noticed that if two electrons could tunnel
simultaneously, this process would become energetically possible at
$eV=\Delta$, and cause the structure in the I-V observed by Taylor and
Burstein~\cite{Taylor1963}. Within this \emph{multiparticle tunneling
theory} the origin of the SGS would be the occurrence of multiple
processes in which $n$ quasiparticles cross simultaneously the contact
barrier. The original perturbative analysis of this theory has serious
problems. In particular, the current was found to diverge at certain
voltage, which avoids to calculate meaningful I-Vs within this approach.
A second explanation was put forward by Werthamer~\cite{Werthamer1966},
who suggested that the SGS could be caused by a self-detection of the ac
Josephson effect. The main problem of this explanation is that it
invokes two different mechanisms for the odd and even terms, while the
experimental current jumps are identical for both series. In 1982
Klapwijk, Blonder and Tinkham~\cite{KBT} introduced the concept of
multiple Andreev reflection (MAR).  In this process a quasiparticle
undergoes a cascade of Andreev reflections in the contact interface (see
Fig.~\ref{MAR}). They showed that a MAR in which a quasiparticle crosses
the interface $n$ times becomes possible at a voltage $eV=2\Delta/n$,
which explains naturally the SGS. The quantitative analysis of the I-Vs
was based on a semiclassical approach which fails away from perfect
transparency~\cite{OBTK,Flensberg1988}.  A few years later, Arnold
reported the first fully microscopic calculation of I-Vs based on a
Green's function approach~\cite{Arnold1987}.

The theoretical discussion was finally clarified with the advent of
modern mesoscopic theories. Using the scattering
formalism~\cite{Bratus1995,Averin1995,Hurd1996} and the so-called Hamiltonian
approach~\cite{Cuevas1996}, different authors reported a complete
analysis of the dc and ac Josepshon effect in point contacts. These
theories clearly showed that the MARs are responsible of the subgap
transport in these systems. They also showed that the multiparticle
tunneling of Schrieffer and Wilkins and the MARs are indeed the same
mechanism. The new microscopic theories have also allowed the
calculation of a series of properties such as resonant 
tunneling~\cite{Yeyati1997,Johansson1999}, shot
noise~\cite{Cuevas1999,Naveh1999} and the Shapiro steps~\cite{Cuevas2002}.

From the experimental point of view, the main problem has always been
the proper characterization of the interface of the superconducting
contact. Uncertainties in the interfaces properties often avoid a
proper comparison between theory and experiment. The situation has
considerably improved with the appearance of the metallic atomic-sized
contacts, which can be produced by means of scanning tunneling
microscope and break-junction techniques~\cite{Muller1992,Post1994,
Rodrigo1994,Koops1996,Scheer1997,Scheer1998,Ludoph2000,Goffman2000,Cron2001}.
These nanowires have turned out to be ideal systems to test the modern
transport theories in mesoscopic superconductors. Thus, for instance
Scheer and coworkers~\cite{Scheer1997} found a quantitative agreement
between the measurements of the current-voltage characteristics of
different atomic contacts and the predictions of the theory for a
single-channel superconducting contact~\cite{Averin1995,Cuevas1996}.
These experiments not only helped to clarify the origin of the SGS, but
also showed that the set of the transmission coefficients in an
atomic-size contact is amenable to measurement. This possibility has
recently allowed a set of experiments that confirm the theoretical
predictions for transport properties such as
supercurrent~\cite{Goffman2000}, noise~\cite{Cron2001} and even
resonant tunneling in the context of carbon nanotubes~\cite{Buitelaar2003}. 
From these combined theoretical and experimental efforts a coherent picture 
of transport in superconducting point contacts has emerged with multiple
Andreev reflections as a central concept.

The most recent development in the understanding of the dc transport in
superconducting contacts is the analysis of the full counting
statistics~\cite{Cuevas2003,Johansson2003}.  Full counting statistics
(FCS) is a familiar concept in quantum optics (see for
instance~\cite{Mandel1995}), which has been recently adapted to electron
transport in mesoscopic conductors by Levitov and
coworkers~\cite{Levitov1993}. FCS gives the probability $P(N)$ that $N$
charge carriers pass through a conductor in the measuring time. Once
these probabilities are known one can easily compute not only the mean
current and noise, but all the cumulants of the current distribution.
Since the introduction of FCS for electronic systems, the theory has
been sophisticated and applied to many different contexts 
(see Ref.~\cite{nazarov:03} for a recent review).

The counting statistics of a one-channel quantum contact has the
surprisingly simple form of a \textit{binomial distribution}
$P(N)=\binom{M}{N} T^N (1-T)^{M-N}$, where $T$ is the transmission
probability and $M\sim V$ is the number of
attempts~\cite{Levitov1993,levitov:96-coherent}. The generalization to
many contacts and/or finite temperatures is straightforward, by noting
that different energies and channels have to be added independently. In
this way, the counting statistics of diffusive contacts at zero
temperature~\cite{levitov:96-diffusive} and at finite temperatures
\cite{Nazarov1999} could be obtained using the universal distribution of
transmission eigenvalues~\cite{dorokhov,nazarov:94-diffusive}. It is
worth to note, that the FCS in the limit of small transparency reduces to
a \textit{Poisson distribution}, which can also be obtained using
classical arguments and neglecting correlations between the different
transfer events. Interestingly, the Poissonian character allows to
directly extract the charge of the elementary event, which can be used
to determine e.g. fractional charges~\cite{fractional,Levitov2001,saleur:01}. 
A general approach to obtain the counting statistics of mesoscopic 
condutors was formulated by Nazarov~\cite{Nazarov1999} using an 
extension of the Keldysh-Green's function method, which allowed to
present the counting statistics of a large class of quantum contacts
in a unified manner~\cite{Belzig2001}. In Ref.~\cite{Cuevas2003} we 
have shown, how this method can be used for a time-dependent transport
problem like a superconducting contact out of equilibrium.

The counting statistics of a contact between a normal metal and a
superconductor at zero temperature and $eV\ll\Delta$ was shown to be
again binomial with the important difference that only even numbers of
charges are transferred~\cite{Muzykantskii:1994}. The probability of an
elementary event is then given by the Andreev reflection coefficient
$R_A=T^2/(2-T)^2$~\cite{beenakker:92}.  Again, the
generalization of this result to many channel conductors is obtained by
summing over independent channels. For a diffusive metal the resulting
statistics was shown to be an exact replica of the corresponding
statistics for normal diffusive transport, provided the double charge
transfer is taken into account \cite{belzig:03-book}. This holds for
coherent transport $eV\ll E_{Th}$, where $E_{Th}$ is the inverse
diffusion time, as well as in the fully incoherent regime $eV\gg E_{Th}$
\cite{belzig:03-incoherent}. For intermediate voltages, correlations of
transmission eigenvalues at different energies modify the distribution
of transmission eigenvalues~\cite{samuelsson:04}, which lead to a
nonuniversal behavior of the transport statistics, predicted
theoretically~\cite{belzig:01-diffusive} and confirmed experimentally
\cite{reulet:03}. Here, we note that a doubling of the noise was
experimentally observed in diffusive wires \cite{2e}, confirming earlier
theoretical predictions~\cite{2e-theory}. However, to trace this back to
a doubling of the elementary charge transfer follows only from an
analysis of the counting statistics. A direct experimental determination
of the doubled charge transfer was recently performed in a conductor
containing a tunnel junction~\cite{lefloch:03}. Here, the
underlying statistics is Poissonian and the noise directly gives access
to the charge of the elementary event~\cite{stenberg:02,pistolesi:03}.

An interesting problem occurs, when one applies the concept of counting
statistics to a supercurrent through a quantum contact
\cite{Belzig2001}. The resulting statistics can not be directly related
to a probability distribution, since some of the 'probabilities' would
be negative. A closer inspection of the formalism showed, that the
interpretation of probabilities relies on the proper definition of a
quantum measuring device~\cite{Kindermann2001,Kindermann2003,romito:04}.
As we will see below, in superconducting contacts out of equilibrium
these problems do not occur and all probabilities are positive.

In Ref.~\cite{Cuevas2003} we have demonstrated that the charge transport
in superconducting point contacts out of equilibrium can be described by
a \textit{multinomial distribution} of processes in which a multiple
charge is transferred.  More importantly, we have shown that the
calculation of the FCS allows us to identify the probability of the
individual MARs and the charge transferred in these processes. This
information probably provides the deepest insight into the transport
properties of these systems. In this sense, in this work we present a
comprehensive analysis of the dc transport properties of superconducting
point contacts from the point of view of the FCS. We shall show that
even in the most well-studied situations the FCS provides a fresh view.
Moreover, we show that the FCS allows a unified description of many
different type contacts. We also extend the analysis presented in
Ref.~\cite{Cuevas2003} to finite temperature.

The paper is organized as follows. In section II, after introducing
some basic concepts of charge statistics, we discuss the calculation of
the cumulant generating functional within the Keldysh-Green's function
approach. Section III is devoted to the calculation of the MAR
probabilities at zero temperature. We present both the results of a toy
model and the full expressions. In section IV we apply the results of
the previous section to describe the different transport properties of
three different situations: (i) a contact between BCS superconductors,
(ii) a contact between superconductor with a modified density of states
due to a pair-breaking mechanisms, and (iii) a short diffusive SNS
contact. In section V we analyze the transport at finite temperature
paying special attention to the third cumulant. Finally, we present our
conclusions in section VI.

\section{Description of the formalism}

\subsection{Some basic concepts}

Our goal is to calculate the full counting statistics of a
superconducting contact. This means that the quantity that we are
interesting in is the probability $P_{t_0}(N)$, that $N$ charges are
transferred through the contact in the time interval $t_0$.
Equivalently, we can find the \emph{cumulant generating function} (CGF)
$S_{t_0}(\chi)$, which is simply the logarithm of the characteristic
function and is defined by

\begin{equation}
\exp(S_{t_0}(\chi)) = \sum_N P_{t_0}(N)\exp(iN\chi)\;.
\label{eq:cgf-def}
\end{equation}
Here, $\chi$ is the so-called counting field. From the knowledge of the
CGF one easily obtains the different cumulants that characterize the
probability distribution
\begin{equation}
  \label{eq:cumulants1}
  C_n=\left.
    \left(-i\right)^n\frac{\partial^n}{\partial \chi^n}
    S_{t_0}(\chi)\right|_{\chi=0}\;.
\end{equation}
Notice that the first cumulants are related to the moments of the
distribution as follows
\begin{eqnarray}
  \label{eq:cumulants}
  C_1  =  \overline N\equiv \sum_N N P_{t_0}(N) & , &
  C_2 = \overline{(N-\overline{N})^2}\,, \nonumber \\
  C_3 = \overline{(N-\overline N)^3}\; ,\;\;
  C_4 & = & \overline{(N-\overline N)^4} - 3 C_2^2\,,
\end{eqnarray}
and so on. It is also important to remark that these cumulants have a
simple relation with the relevant transport properties that are actually
measured. Thus, for instance, the mean current is given by $I = (e/t_0)
C_1$ and the symmetrized zero frequency noise is given by $S_I =
(2e^2/t_0) C_2$\footnote{This relation follows from the definition of
  the current noise power $S_I=2\int_{-\infty}^\infty d\tau \langle
  I(\tau/2)I(-\tau/2)\rangle$. The second cumulant, on the other hand,
  is defined by $C_2=\iint_0^{t_0} dtdt^\prime \langle
  I(t)I(t^\prime)\rangle$. In the static situation the current-current
  correlation function depends only on the time difference
  $\tau=t-t^\prime$ and decays on some characteristic scale $\tau_0$.
  For long observation times $t_0\gg \tau_0$ we find for the second
  cumulant $C_2=(t_0/2e^2) S_I$.}. For higher cumulants such relations
are not straightforwardly obtained, but it can be shown that the
cumulants defined above correspond to the observable quantities in an
electron counting experiment
\cite{Belzig2001,Kindermann2001,Kindermann2003}. Thus, the cumulants
represent all information, which is available in a measurement of the
charge accumulated during the observation period $t_0$.

\subsection{Keldysh-Green's function approach to FCS}

As mentioned above, our system of interest is a voltage-biased
superconducting point contact, i.e. two superconducting electrodes
linked by a constriction, which is much shorter than the superconducting
coherence length. We concentrate ourselves on the case of a single
channel contact described by a transmission probability $T$. The main
difficulty in the determination of the FCS arises from the ac-Josephson
effect. Here, a constant applied bias voltage $eV$ gives rise to
time-dependent currents as a consequence of the Josephson relation
$(\partial/\partial t)\phi(t)=2eV/\hbar$. In the long-time limit $t_0\gg
\hbar/eV$ these oscillating currents do not contribute to the net charge
transfer, in which we are interested. However, this intrinsic
time-dependence is reflected in the CGF and a little care has to be
taken, when the FCS is defined.

To obtain the FCS in a superconducting point contact we make use of the
Keldysh-Green's function approach to FCS introduced by Nazarov and one
of the authors~\cite{Nazarov1999,Belzig2001}, and we refer to reader to
these papers for further details on the basis of this theoretical
approach. In what follows, we concentrate ourselves on the specific
difficulties introduced in the case of a contact between two
superconductors. Our starting point for the determination of the CGF is
to define the relation between the CGF and the counting current in
analogy to Refs.~\cite{Nazarov1999,Belzig2001}:

\begin{equation}
 \label{eq:chi-current}
 \frac{\partial }{\partial \chi} S_{t_0}(\chi) = \frac{i}{e}
 \int_0^{t_0} dt I(\chi,t)\,.
\end{equation}
This scalar current can be calculated in terms of the \emph{matrix
  current} which describes the transport properties of the contacts.
Nazarov has shown that, in the case of short junctions the matrix
current (in Keldysh-Nambu space) adopts the following
form~\cite{Nazarov1999b}

\begin{equation}
  \label{eq:matrix-current}
  \check{I}(\chi,t,t^{\prime}) = -\frac{e^2}{\pi}
      \left(\frac{2T\left[\check G_1(\chi)\stackrel{\otimes}{,} 
      \check G_2\right]}
    {4+T\left(\{\check G_1(\chi)\stackrel{\otimes}{,}\check
        G_2\}-2\right)}\right) (t,t^\prime)\,.
\end{equation}
Here $\check G_{1(2)}(t,t^\prime)$ denote the matrix Green's functions
on the left and the right of the contact. In our problem these functions
depend on two time arguments and the products $\otimes$ appearing in
Eq.~(\ref{eq:matrix-current}) should be understood as convolutions over
the intermediate time arguments, i.e. $(A\otimes B)(t,t^\prime)=\int
dt^{\prime\prime} A(t,t^{\prime\prime})B(t^{\prime\prime},t^\prime)$.
It is worthwhile to note, that the derivation for the matrix current in
Ref.~\cite{Nazarov1999b} was done for Green's functions in the static
situation, in which case all Green's functions depend only on
$t-t^\prime$. However, the derivation can be directly taken over to
time-dependent problems, because the time-dependent Green's functions
satisfy the normalization condition
\begin{equation}
  (\check G\otimes \check G)(t,t^\prime) = 
  \check\delta(t-t^\prime)\,.
  \label{eq:normalization}
\end{equation}
Finally, the time-dependent scalar current is obtained from the matrix
current by
\begin{equation}
  \label{eq:el-current}
  I(\chi,t) = \frac{1}{4e} {\rm Tr}\left[ \check\tau_K\check
  I(\chi,t,t)\right] \,, 
\end{equation}
where $\check\tau_{\rm  K}=\hat\sigma_3\bar\tau_3$ is a matrix in 
Keldysh($\hat{\ }$)-Nambu($\bar{\ }$) space. $\hat\sigma_i(\bar\tau_i)$
are the standard Pauli matrices in Keldysh(Nambu)-space.

Let us now describe Green's functions entering
Eq.~(\ref{eq:matrix-current}).  The counting field $\chi$ is
incorporated into the matrix Green's function of the left electrode as
follows
\begin{equation}
\label{eq:countrot}
\check G_1(\chi,t,t^{\prime}) = e^{-i\chi\check \tau_K/2}
\check G_1(t,t^{\prime}) e^{i\chi\check \tau_K/2}\,.
\end{equation}
Here $\check G_1(t,t^{\prime})$ is the reservoir Green's function in the
absence of the counting field. We set the chemical potential of the
right electrode to zero and represent the Green's functions by
\begin{equation}
  \check G_1(t,t^{\prime}) = e^{i \phi(t) \bar \tau_3/2} \check G_S(t-t^{\prime})
  e^{-i \phi(t^\prime) \bar \tau_3/2}
  \label{eq:phaserot}
\end{equation}
and $\check G_2(t,t^{\prime}) = \check G_S(t-t^{\prime})$. Here,
$\phi(t) = \phi_0 + (2eV/\hbar) t$ is the time-dependent superconducting
phase difference, and $\phi_0$ is its dc part. $\check G_S$ is the
Green's function of a superconducting reservoir (we consider the case of
a symmetric junction), which reads
\begin{eqnarray}
  \label{eq:reservoir}
  \check G_S(t-t^\prime) & = &
  \int dE \; G_S(E)e^{iE(t-t^\prime)}\,, \\\nonumber
  \check G_S(E) &= & \left( \begin{array}[c]{cc}
  (\bar A - \bar R) f + \bar R & (\bar A - \bar R) f \\
  (\bar A - \bar R) (1 - f) & (\bar R - \bar A) f + \bar A
  \end{array}\right).
\end{eqnarray}
Here, $\bar R(\bar A)(E)$ are retarded and advanced Green's functions of
the leads and $f(E)$ is the Fermi function. Advanced and retarded
functions in (\ref{eq:reservoir}) have the Nambu-structure $\bar R(\bar
A) = g^{\text{R,A}}\bar\tau_3 + f^{\text{R,A}}\bar\tau_1$ fulfilling the
normalization condition $f^2+g^2=1$. They depend on energy and the
superconducting order parameter $\Delta$.

Using the time dependence of the leads Green's functions it is easy to
show from Eq.~(\ref{eq:matrix-current}) that the scalar current admits
the following Fourier series

\begin{equation}
\label{eq:harmonics}
I(\chi,t) = \sum_n I_n(\chi) e^{i n \phi(t)} \;,
\end{equation}

\noindent
which means that the current oscillates with all the harmonics of the
Josephson frequency. It is important to stress that the components
$I_n(\chi)$ are independent of dc part of the superconducting phase. In
this work we only want to consider the dc part of the CGF.  For this
purpose, we take the limit of a long measuring time $t_0$, much larger
than the inverse of the Josepshon frequency, and hereafter we drop the
subindex $t_0$ in the expression of the CGF. From Eq.~(\ref{eq:chi-current})
and Eq.~(\ref{eq:harmonics}) it is obvious that by selecting the dc 
component, the dc part of the phase drops the calculation and the CGF
is free of the problems related to gauge invariance found for the dc 
Josephson effect~\cite{Belzig2001,Kindermann2001,Shelankov:03}.

Keeping in mind the presence of the time integration described aboved,
and with the help of Eqs.~(\ref{eq:matrix-current}-\ref{eq:el-current}),
one can integrate Eq.~(\ref{eq:chi-current}) to obtain the following
expression for the CGF of superconducting
constrictions~\cite{Belzig2001}
\begin{equation}
  \label{eq:cgf}
  S(\chi) = \frac{t_0}{h} {\textrm{Tr}}
  \ln\left[1+\frac{T}{4}
  \left(\{\check G_1(\chi), \check G_2\}_\otimes -2\right)\right]\;.
\end{equation}
The symbol $\otimes$ implies that the products of the Green's functions
are convolutions over the internal energy arguments, i.~e.
\begin{equation}
  (G_1 \otimes G_2) (E,E^{\prime}) = \int dE_1 \; G_1(E,E_1)
  G_2(E_1,E^{\prime})\,.
\end{equation}
The trace runs not only over the Keldysh-Nambu space, but also includes
a trace in the energy arguments, i.e. $\int dE \; g(E,E)$. 

The time-dependent Green's functions of Eq.~(\ref{eq:countrot}) fulfill 
the normalization condition of Eq.~(\ref{eq:normalization}). This enables
us to use the relation 
\begin{equation}
  2-\{\check G_1,\check G_2\}_\otimes = \left( \check G_1 -
    \check G_2 \right)^2_\otimes
\end{equation}
to write the CGF as 
\begin{equation}
  S(\chi) = \frac{t_0}{h}
  {\textrm{Tr}} \left\{ \ln \check Q_+ + \ln \check Q_- \right\}\,,
\end{equation}
where $\check Q_{\pm} \equiv 1 \pm (\sqrt{T}/2) \left( \check G_1(\chi)
  - \check G_2 \right)$.  One can show that both logarithms give the
same contribution, and therefore we concentrate in the analysis of the
first one, and we drop the subindex $+$. Additionally, we use the
relation $\textrm{Tr}\ln\check Q= \ln \det \check Q$ to write the CGF as

\begin{equation}
  \label{eq:CGF-simplified}
  S(\chi) = \frac{t_0}{h} \ln \det \check Q(\chi) .
\end{equation}

Thus, at this stage the calculation reduces to the calculation of the
determinant of a infinite matrix. Due to the time dependence of the
lead Green's functions their form in energy space is 
$\check G(E,E^{\prime}) = \sum_{n} \check G_{0,n}(E)
\delta(E - E^{\prime} +neV)$, where $n=0,\pm 2$. This implies that
the matrix $\check Q$ also admits the same type of representation,
which in practice means that $\check Q$ is a block-tridiagonal
matrix of the form

\begin{equation}
\check Q = \left(
\begin{array}[c]{ccccccc}
\ddots & \ddots & \ddots &\check 0\\
& \check Q_{-2,-4} & \check Q_{-2,-2} & \check Q_{-2,0} & \check 0 \\
&  \check 0 &  \check Q_{-2,0} & \check Q_{0,0} & \check Q_{0,2} & \check 0 \\
& & \check 0 & \check Q_{2,0} & \check Q_{2,2} & \check Q_{2,4} \\
&  & & \check 0 & \ddots  & \ddots & \ddots
\end{array}\right)\,, \nonumber
\end{equation}

\noindent
where we have used the notation $\check Q_{n,m} = \check Q(E+neV,E+meV)$.
The different $(4\times 4)$ matrices $\check Q_{n,m}$ have the following
explicit form in terms of the advanced and retarded Green's functions
$g^{R,A}$ and $f^{R,A}$ (remember that we consider a symmetric junction)

\begin{widetext}
\begin{eqnarray}
\check Q_{n,n} & = & \check 1 + \frac{\sqrt{T}}{2} \left(
\begin{array}[c]{cccc}
\rho_{n+1} - \rho_n + g^R_{n+1} - g^R_n &
- \tilde \rho_n - f^R_n &
e^{-i\chi} \rho_{n+1} - \rho_n &
-\tilde \rho_n
\\  
- \tilde \rho_n - f^R_n & 
\rho_n -\rho_{n-1} g^R_n - g^R_{n-1} &
- \tilde \rho_n &
-e^{i\chi} \rho_{n-1} + \rho_n 
\\
e^{i\chi} \delta_{n+1} - \delta_n &
-\tilde \delta_n &
\rho_n - \rho_{n+1} + g^A_{n+1} - g^A_n &
-f^A_n + \tilde \rho_n
\\
-\tilde \delta_n &
-e^{-i\chi} \delta_{n-1} + \delta_n &
-f^A_n + \tilde \rho_n &
\rho_{n-1} - \rho_n + g^A_n - g^A_{n-1}
\\
\end{array}\right) \nonumber \\
& & \check Q_{n,n+2} = \frac{\sqrt{T}}{2} \left(
\begin{array}[c]{cccc}
0 & e^{-i\chi} (\tilde \rho_{n+1} + f^R_{n+1}) & 0 & \tilde \rho_{n+1} \\
0 & 0 & 0 & 0 \\
0 & \tilde \delta_{n+1} & 0 & e^{i\chi} (f^A_{n+1} - \tilde \rho_{n+1}) \\
0 & 0 & 0 & 0 \\
\end{array}\right) \nonumber \\
& & \check Q_{n,n-2} = \frac{\sqrt{T}}{2} \left(
\begin{array}[c]{cccc}
0 & 0 & 0 & 0 \\
e^{i\chi} (\tilde \rho_{n-1} + f^R_{n-1}) & 0 & \tilde \rho_{n-1} & 0 \\
0 & 0 & 0 & 0 \\
\tilde \delta_{n-1} & 0 & e^{-i\chi} (f^A_{n-1} - \tilde \rho_{n-1}) & 0 \\
\end{array}\right)  \;,
\label{eq:Q-matrices}
\end{eqnarray}
\end{widetext}

\noindent
where we have used the shorthand notation $g^{R,A}_n = g^{R,A}(E+neV)$,
$\rho = (g^A-g^R) f$, $f$ being the Fermi function, $\tilde \rho =
(f^A-f^R) f$, $\delta = (g^A-g^R) (1-f)$, and $\tilde \delta = (f^A-f^R)
(1-f)$.

One can restrict the fundamental energy interval to $E-E^\prime \in [0,eV]$,
and therefore the CGF adopts the form
$S(\chi) = (t_0/h) \int^{eV}_0 dE\; \ln \det \check Q$.
From Eq.~(\ref{eq:Q-matrices}), it is obvious that $\det \check Q$ can
be written as the following Fourier series in $\chi$

\begin{equation}
\det \check Q(\chi) = \sum^{n=\infty}_{n=-\infty} P^{\prime}_n(E,V) e^{in \chi} \;,
\end{equation}

\noindent
where the coefficients $P^{\prime}_n(E,V)$ have still to be determined.
Keeping in mind the normalization $S(0) = 0$, it is clear that one can
rewrite the CGF in the following form

\begin{equation}
  \label{eq:marfcs}
  S(\chi) = \frac{t_0}{h}\int_0^{eV} dE \ln \left[
  1+\sum_{n=-\infty}^{\infty}P_n(E,V)\left(e^{in\chi}-1\right)\right]\,,
\end{equation}

\noindent
where 

\begin{equation}
P_n(E,V) = P^{\prime}_n(E,V) / \sum^{n=\infty}_{n=-\infty} P^{\prime}_n(E,V) .
\end{equation}

\noindent
Eq.~(\ref{eq:marfcs}) has the form of the CGF of a multinomial
distribution in energy space (provided more than one $P_n$ is different
from zero).  The different terms in the sum in Eq.~(\ref{eq:marfcs})
correspond to transfers of multiple charge quanta $ne$ at energy $E$
with the probability $P_n(E,V)$, which can be seen by the
$(2\pi/n)$-periodicity of the accompanying $\chi$-dependent counting
factor. This is the main result of our work and it proves, that the
charges are indeed transferred in large quanta.  Of course, we still
have to determine the probabilities $P_n(E,V)$, which is a non-trivial
task and it will the goal of the next section.

\subsection{Cumulants}
\label{sec:cumulants}

As explained before, from the CGF one can easily calculate the
cumulants of the distribution and in turn many transport properties.  Of
special interest are the first three cumulants $C_1$, $C_2$ and $C_3$,
which correspond to the average, width and skewness of the distribution
of transmitted charge, respectively. From Eq.~(\ref{eq:cumulants1}) and
Eq.~(\ref{eq:marfcs}), it follows that these cumulants can be expressed
in terms of the probabilities $P_n(E,V)$ as follows
\begin{widetext}
\begin{eqnarray}
  \label{eq:cumulants2}
  C_1 & = & \frac{t_0}{h} \int^{eV}_0 dE \; \sum_n n P_n\,, \\
  C_2 & = & \frac{t_0}{h} \int^{eV}_0 dE \; \left\{ \sum_n n^2 P_n -
    \left( \sum_n n P_n \right)^2 \right\}\,, \\
  C_3 & = & \frac{t_0}{h} \int^{eV}_0 dE \; \left\{ \sum_n n^3 P_n 
    + 2 \left(\sum_n n P_n \right)^3 
    - 3 \left( \sum_n n P_n \right) \left( \sum_n n^2 P_n \right)  \right\} \;.
\end{eqnarray}
\end{widetext}
These expressions are a simple consequence of the fact that the charge
transfer distribution is multinomial in energy space. At
zero temperature the sums over $n$ are restricted to positive values $(n
\ge 1)$. We remind the reader that the first two cumulants are
simply related to the dc current, $I = (e/t_0) C_1$, and to the
zero-frequency noise $S_I =(2e^2/t_0) C_2$.

It is instructive to discuss some consequences of these expressions. Let
us first recall, what happens when only one process contributes, which
has, \textit{e.g.}, the order $n$. The first three cumulants are
\begin{eqnarray}
  \label{eq:c1n}
  C_{1;n} & = &  n \int^{eV}_0 \frac{t_0dE}{h} P_n\,, \\
  \label{eq:c2n}
  C_{2;n} & = & n^2 \int^{eV}_0 \frac{t_0dE}{h} P_n \left( 1 - P_n \right)\,, \\
  \label{eq:c3n}
  C_{3;n} & = & n^3 \int^{eV}_0 \frac{t_0dE}{h} P_n \left( 1 - P_n \right)
  \left(1-2P_n\right)\,.
\end{eqnarray}
We see, that the $i^\mathrm{th}$ cumulant is proportional $n^i$, i.~e.
the $i^\mathrm{th}$ power of the charge of the respective elementary
event. The expressions under the integral in
Eqs.~(\ref{eq:c1n}-\ref{eq:c3n}) have the same form as for binomial
statistics, however in general the $P_n(E,V)$ depend on energy in a
nontrivial way and the energy-integrated expressions for the cumulants
do not correspond to binomial statistics. A simple interpretation in
terms of an effective charge transferred is only possible if
$P_n(E,V)\ll 1$ for \textit{all} energies, in which case one recovers
the standard result for Poisson statistics, $C_{i;n}=n^{i-1}C_{1;n}$.
According to Eq.~(\ref{eq:c3n}) the sign of the spectral third cumulant
can be positive or negative, depending on the size of $P_n$ (positive
for $P_n<1/2$ and negative for $P_n>1/2$). The overall sign depends on
the energy average and is not simple to predict. Note, however, that the
probabilities of MAR-processes of higher orders decrease approximately
as $~T^n$. We may therefore speculate that to obtain a negative third
cumulant for higher order processes we will need more open contacts (a
rough estimate is thus that $T\gtrsim 1/\sqrt[n]{2}$ to have $P_n\gtrsim
1/2$ and, therefore, $C_3<0$).

The general statistics (\ref{eq:marfcs}) is a \textit{multinomial}
distribution and it is therefore interesting to compare with
\textit{independent} binomial distributions. This is most easily done by
assuming, that only two processes compete. Taking these processes to be
of order $n$ and $m$ the first three cumulants read
\begin{widetext}
\begin{eqnarray}
  \label{eq:c1nm}
  C_{1;nm} & = & C_{1;n}+C_{1;m}\,, \\
  \label{eq:c2nm}
  C_{2;nm} & = & C_{2;n}+C_{2;m}- 2nm \int^{eV}_0 \frac{t_0dE}{h} P_n P_m \,, \\
  \label{eq:c3nm}
  C_{3;nm} & = & C_{3;n}+C_{3;m}
  - 3nm \int^{eV}_0 \frac{t_0dE}{h} P_n
  P_m \left[ n(1 - P_n)+m(1-P_m) \right]\,.
\end{eqnarray}
\end{widetext}
We see that the first cumulant is just the sum of the contributions of
the different processes $n$ and $m$ and we therefore have to look at
higher cumulants to gain information on correlations between the
processes of different order. In both, the second and the third
cumulant, such correlations appear and it is evident from
Eqs.~(\ref{eq:c2nm}) and (\ref{eq:c3nm}) that both are
reduced below the value obtained for independent binomials. The
correlation terms appear inside the energy integration and therefore
both processes must be possible at the same energy.

Finally, we note that in order to study correlation between $N$
different processes one would have to look at the $N$th order
cumulant. This becomes clear if one notices that only the $N$th cumulant
contains a term with products of $N$ probabilities and therefore the
possibility to have a product of probabilities of $N$ different
processes. 

\section{MAR probabilities: zero temperature}

This section is devoted to the calculation of the probabilities $P_n(E,V)$
at zero temperature. First, we discuss a simple model which nicely illustrates 
the transmission dependence of these probabilities, and secondly we present
the general expressions.

\subsection{Toy model}

To obtain a feeling for the forthcoming calculations we will now study a
strongly simplified model of a superconducting contact. For that
purpose, let us assume that we can neglect the Andreev reflections for
energies outside the gap region and replace the quasiparticle density of
states by a constant for $|E|>\Delta$. Furthermore, we neglect that
energy-dependent phase shift $\sim\textrm{acos}(E/\Delta)$, usually
associated with the finite penetration of excitations close to the gap
edge.  Mathematically, this means that we set $f^{R,A}(|E|<\Delta)=1$,
$g^{R(A)}(|E|>\Delta)=\pm 1$, and both are equal to zero otherwise. This
simplifies the calculation a lot, since the matrix $\check Q$ in
Eq.~(\ref{eq:CGF-simplified}) now becomes finite. In particular, for
subharmonic voltage $eV=2\Delta/n$ the matrix is also
energy-independent. It is interesting to note that the toy model is also
able to describe the counting statistics of normal contacts and Andreev
contacts.

To facilitate the discussion of the matrix structure it is useful to
introduce the 2$\otimes$2 matrix in the Keldysh subspace
\begin{eqnarray}
  \hat K_\pm(\chi) & = &
  \mp \hat \tau_3 -2\hat \tau_\pm e^{\pm i\chi}\,,
\end{eqnarray}
where $\hat\tau_i$ are Pauli matrices and
$\hat\tau_{\pm}=\left(\hat\tau_1\pm i\hat\tau_2\right)/2$. In fact, $\hat
K_\pm$ correspond to occupied(empty) quasiparticle states (for
$E>|\Delta|$). The matrix structure for superconducting or normal
terminals is summarized in Table~\ref{tab:toy1}.
\begin{table}[bt]
\[
\begin{array}{r||c|c|c|}
  \textrm{super:}& E>\Delta & |\Delta|>E & -\Delta>E\\\hline
  \textrm{normal:}  & E>eV & & E<eV\\\hline\hline
  \hat g_{11}(\chi)  & \hat K_-(\chi) &  0 & \hat K_+(\chi) \\\hline
  \hat g_{22}(\chi)  & -\hat K_-(-\chi) & 0 & -\hat K_+(-\chi) \\\hline
  \hat g_{12(21)}(\chi)  & 0 & e^{\pm i\chi\check\tau_3} & 0\\\hline
\end{array}
\]
\caption{Green's functions in the toy model. The indices $\hat
  g_{\alpha\beta}$ denote the respective element in Nambu space. $\hat K_{\pm}
  = \mp \hat \tau_3 -2\hat \tau_\pm e^{\pm i\chi}$ denotes a matrix in
  Keldysh space. The table holds for left and right terminal, provided
  the energies and the counting fields are chosen properly.}
\label{tab:toy1}
\end{table}
The counting statistics is obtained from the general relation (\ref{eq:CGF-simplified})
\begin{equation}
  S(\chi,V) =\frac{2t_0}{h} \int_0^{eV} dE \; \textrm{Tr} 
  \ln\left[1+\frac{\sqrt T}{2}\left(\check G_1(\chi)-\check G_2\right)\right]\,.
\end{equation} 
The incorporation of the energy discretization is obtained by a
redefinition of the trace in the above formulas, and a limitation of
the energy integration to an interval of width $eV$. Note, that we
have to evaluate only one of the two terms $Q_\pm$, since the FCS can only
depend on $T$ and not on $\sqrt T$.

To calculate the determinant we note that $\check Q$ is a band matrix
of width 3 in the energy index. Then the following reduction formula
for the determinant is useful (assuming a block starts at some $n$,
which we arbitrarily set to zero):
\begin{equation}
  \begin{array}[c]{l}
    \left|
    \begin{array}[c]{cccc}
      \check Q_{0,0} & \check Q_{0,2} & 0 & 0\\
      \check Q_{2,0} & \check Q_{2,2} & \check Q_{2,4} & 0\\
      0 &  \check Q_{4,2} & \check Q_{4,4} & \ddots \\
      0 & 0 & \ddots & \ddots
    \end{array}
  \right| = \\
  \left|\check Q_{0,0}\right|
  \left|
    \begin{array}[c]{ccc}
      \check Q_{2,2}- \check Q_{2,0}\check Q_{0,0}^{-1}\check Q_{0,2} & 
      \check Q_{2,4} & 0 \\
      \check Q_{4,2} & \check Q_{4,4} & \ddots\\
      0 & \ddots &\ddots
    \end{array}
  \right|\,.
\end{array}
\end{equation}
Another useful property (which holds in the toy model) is the Nambu
structure of the $\check Q$'s, see Eq.~(\ref{eq:Q-matrices}) and
Table~\ref{tab:toy1}: diagonal components in energy space, i.~e. $\check
Q_{n,n}$, are always block-diagonal in Nambu space and the off-diagonal
components $Q_{n,n\pm 2}$ are purely off-diagonal in Nambu space and
diagonal in Keldysh-space. Consequently, $\check Q_{n-2,n-2}-
Q_{n-2,n}Q_{n,n}^{-1}Q_{n,n-2}$ appearing in the expansion of the
determinant is block-diagonal again and the whole calculation of the
determinant (\ref{eq:CGF-simplified}) boils down to a recursive
calculation of determinants and inversions of 2$\times$2-matrices.  This
will become more clear, when we will treat the explicit examples below.

\subsubsection{Normal Contact}

It is instructive to demonstrate the procedure first for a normal
contact. The Green's functions are (we restrict the calculation here to
electron block, the hole block gives actually the same contribution)
\begin{equation}
  \hat G_1 = \left\{
    \begin{array}[c]{lcl}
      \hat K_-(\chi) & , & n\ge 0\\
      \hat K_+(\chi) & , & n<0
    \end{array}\right.\,,\,
  \hat G_2 = \left\{
    \begin{array}[c]{lcl}
      \hat K_-(0) & , & n> 0\\
      \hat K_+(0) & , & n\le0
    \end{array}\right. \;.
\end{equation}
Note that we have chosen the fundamental energy interval $[-eV/2,eV/2]$,
since then the Green's functions are constant inside each interval.  Then
we find
\begin{equation}
  \label{eq:qn}
  \frac{(\hat Q-1)_{n,m}}{\sqrt T/2} = \delta_{n,m}\left\{
    \begin{array}[c]{ll}
      \hat\tau_+(e^{i\chi}-1) & n>0\\
      \hat\tau_3+\hat\tau_+e^{i\chi}
      -\hat\tau_-  & n=0\\
      \hat\tau_- (e^{-i\chi}-1) & n<0
    \end{array}\right. \;.
\end{equation}
The matrix $Q$ has thus block diagonal form. The blocks $n>0$ and $n<0$
are tridiagonal and the determinants are all equal to 1. The remaining
determinant of the $n=0$ block is
\begin{equation}
  \textrm{det}
  \begin{pmatrix}
    1+\sqrt T & \sqrt T e^{i\chi} \\
    -\sqrt T & 1-\sqrt T
  \end{pmatrix}
  =1-T+Te^{i\chi}\,.
\end{equation}
The CGF is finally $S(\chi)=(2eVt_0/h)\ln(1+T(e^{i\chi}-1))$ in agreement
with Levitov and Lesovik \cite{Levitov1993}. Notice that a factor of 2
enters the CGF, because we get an additional contribution from the
hole block (thus it is due to spin).

\subsubsection{Andreev contact}

We now consider a contact in which one of the sides is superconducting
and the other is a normal metal. Again, the calculation can be done in a
similar way.  Here we apply a voltage $|eV|\ll\Delta$ to the normal
contact. The Green's functions are again diagonal in the energy space,
since we assume that the superconductor is at zero potential. For the
normal metal we find (taking as fundamental energy interval $[-eV,eV]$)
\begin{eqnarray}
  \left(\hat G_1\right)_{11} & = & 
  \left\{
    \begin{array}[c]{lcl}
      \hat K_-(\chi) & , &  n\ge 0\\
      \hat K_+(\chi) & , & n<0
    \end{array}\right.\,,\nonumber\\
  \left(\hat G_1\right)_{22} & = & \left\{
    \begin{array}[c]{lcl}
      -\hat K_-(\chi) & , & n>0\\
      -\hat K_+(\chi) & , & n\le 0
    \end{array}\right.
\end{eqnarray}
and for the superconductor $\left(\hat G_2\right)_{12}=\left(\hat G_2\right)_{21}=\hat\tau_1$ and 0
otherwise. The only non-zero block is the $n=0$ energy block
\begin{equation}
  \label{eq:andreevblock}
    \check G_1(\chi)-\check G_2 =
  \begin{pmatrix}
    K_-(\chi) & -1\\
    -1 & -K_+(-\chi)
  \end{pmatrix}\,,
\end{equation}
which yields for the CGF in the form (\ref{eq:cgf}) the determinant of
\begin{equation} 
  \check Q=
  \begin{pmatrix}
    1-\frac{T}{2} & \frac{T}{4}(\hat K_+-\hat K_-)\\
    \frac{T}{4}(\hat K_+-\hat K_-) & 1-\frac{T}{2}
  \end{pmatrix}\,.
\end{equation}
To calculate the determinant we subtract from rows 3 and 4 the
rows 1 and 2 multiplied with $\frac{T}{4}(1-\frac{T}{2})(\hat
K_+ - \hat K_-)$ and make use of the fact that $(\hat K_--\hat
K_+)^2=4(1-e^{i2\chi})$. The matrix is then tridiagonal and its
determinant is
\begin{equation}
  \left(1-\frac{T}{2}\right)^2
  \left[1+\frac{T^2}{(2-T)^2}\left(e^{i2\chi}-1\right)\right]\,.
\end{equation}
The prefactor is canceled because we are operating under the ln and
have to normalize. Notice that the evaluation of the determinant outside
the transport window can be done in a similar way. One obtains for the
determinant of one block $(1-T/2)^2-T^2(\hat K_-(\chi)-\hat
K_-(-\chi))^2=(1-T/2)^2$, which is independent of the counting field
$\chi$ and is therefore canceled after normalization of the CGF.
Finally we obtain for the FCS (collecting all prefactors)
\cite{Muzykantskii:1994}
\begin{equation}
  \label{eq:fcs-andreev}
  S(\chi)=\frac{2 eVt_0}{h}\ln \left[1+\frac{T^2}{(2-T)^2}\left(e^{i2\chi}-1\right)\right]\,.
\end{equation}
The statistics corresponds to a binomial distribution of charge
transfers. The Andreev reflection leads to a $\pi$-periodicity in $\chi$
which shows that only couples of charges can be transferred and the
charge transfer probability for odd charge numbers vanishes. The number
of attempts, determined by the prefactor of the ln in
(\ref{eq:fcs-andreev}), remains unchanged in comparison to the normal
case.

\subsubsection{Superconducting point contact}

We now come to the main subject of the article, a point contact between
two superconducting banks held at different chemical potentials.  To
write down the general matrix structure of the FCS in the toy model, let
us first obtain the condition for energies to be subgap. Here, we
restrict ourselves to subharmonic voltages, which we write in general as
$eV=2\Delta/(N-1)$, where $N$ denotes the order. The dominating charge
transport mechanism we expect is that $N$ charges are transferred. In
the toymodel, it is the only transport mechanism (since Andreev
reflections above the gap are neglected). To obtain a single-valued
matrix entries, it is favourable to choose as fundamental energy
interval $[0,eV]$ for even $N=2M$ and $[-eV/2,eV/2]$ for odd $N=2M-1$.
For the Nambu row indices of the Green's function of the left terminal
we find
\begin{equation}
  \label{eq:row}
  \begin{array}[c]{r|c|c}
    \textrm{Nambu} & \textrm{Order} & |E|\leq \Delta\\\hline
    \textrm{upper} & \textrm{odd} & -M\leq n\leq M-1 \\
    \textrm{lower} & \textrm{odd} & -M+1\leq n\leq M \\
    \textrm{upper} & \textrm{even} & -M-1\leq n\leq M-1\\
    \textrm{lower} & \textrm{even} & -M\leq n\leq M 
  \end{array}
\end{equation}
The row indices in Nambu space of the right Green's functions have the
energy arguments of upper and lower row interchanged.

To clarify the matrix structure we have prepared a small table. Each
entry denotes the energy for the structure $\left(
  \begin{array}[c]{c|c}
    \hat g_{1i}^L & \hat g_{1i}^R \\\hline \hat g_{2i}^L & \hat g_{2i}^R 
  \end{array}\right)$, where the second (Nambu-) index $i=1,2$ plays no
role. The entries are denoted by $+$ for $E>\Delta$, $0$ for
$|E|\leq\Delta$, and $-$ for $E<-\Delta$.
\begin{equation}
  \begin{array}[c]{r||c|c|c|c|c}
    n & N=2 & N=3 & N=4 & N=5 & N=6\\\hline\hline
    2  & \nmat{+}{+}{+}{+} & \nmat{+}{+}{+}{+} & \nmat{+}{+}{+}{+} 
    & \nmat{+}{0}{0}{+}  & \nmat{+}{0}{0}{+}
    \\[3mm]\hline
    1  & \nmat{+}{+}{+}{+} & \nmat{+}{0}{0}{+} & \nmat{+}{0}{0}{+} 
    & \nmat{0}{0}{0}{0} & \nmat{0}{0}{0}{0}
    \\[3mm]\hline
    0  & \nmat{+}{0}{0}{+} & \nmat{0}{0}{0}{0} & \nmat{0}{0}{0}{0} 
    & \nmat{0}{0}{0}{0} & \nmat{0}{0}{0}{0} 
    \\[3mm]\hline
    -1 & \nmat{0}{-}{-}{0} & \nmat{0}{-}{-}{0} & \nmat{0}{0}{0}{0} 
    & \nmat{0}{0}{0}{0} & \nmat{0}{0}{0}{0} 
    \\[3mm]\hline
    -2 & \nmat{-}{-}{-}{-} & \nmat{-}{-}{-}{-} & \nmat{0}{-}{-}{0} 
    & \nmat{0}{-}{-}{0} & \nmat{0}{0}{0}{0} 
    \\[3mm]\hline
    -3 & \nmat{-}{-}{-}{-} & \nmat{-}{-}{-}{-} & \nmat{-}{-}{-}{-} 
    & \nmat{-}{-}{-}{-} & \nmat{0}{-}{-}{0}
\end{array}
\end{equation}
We observe that the matrix structure in all cases is similar.  A block
with $0$ and $+$ elements, i.e. connecting the quasiparticle states
above the gap to the subgap region is followed a number of blocks inside
the gap (depending on the applied voltage and, finally, is connected by
a block with $0$ and $-$ elements to quasiparticle states below the gap.

Let us now discuss the case $N=2$ ($eV=2\Delta$). Here the relevant
$8\times 8$-matrix is
\begin{equation}
  \frac{\check Q-1}{\sqrt T/2} =
  \left(
    \begin{array}[c]{cccc}
      \hat K_-(\chi) & 0 & 0  & -1\\
      0 & \hat K_-(0) & e^{-i\chi\hat\tau_3} & 0 \\
      0 & e^{i\chi\hat\tau_3} & -\hat K_+(0) & 0 \\
      -1 & 0 & 0 & -\hat K_+(-\chi)
    \end{array}
  \right)
\end{equation}
We observe, that the matrix decouples into two blocks of $4\times 4$
matrices
\begin{equation}
  \check Q_{2A} =1+\frac{\sqrt{T}}{2} 
  \left(
    \begin{array}[c]{cc}
      \hat K_-(\chi) & -1 \\
      -1 & -\hat K_+(-\chi) 
    \end{array}
  \right)
\end{equation}
and
\begin{equation}
  \check Q_{2B} = 1+\frac{\sqrt{T}}{2} 
  \left(
    \begin{array}[c]{cc}
      \hat K_-(0) & e^{-i\chi\hat\tau_3} \\
      e^{i\chi\hat\tau_3} & -\hat K_+(0)
    \end{array}
  \right)\,.
\end{equation}
By comparison with Eq.~(\ref{eq:andreevblock}) we see that
$\ln\textrm{det}\check Q_{2A}$ yields the counting statistics of usual
Andreev reflection. $\check Q_{2B}$ gives actually the same result. This
is most easily seen, if the unitary transformation $\check U\check
Q_{2B}\check U^{\dagger}$ with $\check
U=\textrm{diag}(e^{i\tau_3\chi/2},e^{-i\tau_3\chi/2})$ is applied, which
transforms $\check Q_{2B}$ into $\check Q_{2A}$. Note, that the signs of
the off-diagonal matrices do not matter, since they can be eliminated by
similar unitary transformations. The counting statistics
is therefore given by Eq.~(\ref{eq:fcs-andreev}), the same as for the
Andreev contact.

Now we come to the slightly more complicated case $N=3$
($eV=2\Delta/2$). Here we encounter the matrix
\begin{equation}
  \left(\begin{array}[c]{cccccc}
      \hat K_-(\chi) & 0 & 0 & -1 & 0 & 0\\
      0 & \hat K_-(0) & e^{-i\chi\hat\tau_3} & 0 & 0 & 0
      \\
      0 & e^{i\chi\hat\tau_3} & 0 & 0 & 0 & -1\\
      -1 & 0 & 0 & 0 & e^{-i\chi\tau_3} & 0\\
      0 & 0 & 0 & e^{i\chi\hat\tau_3} & -\hat K_+(0) & 0\\
      0 & 0 & -1 & 0 & 0 & -\hat K_+(-\chi)
  \end{array}\right) \;.
\end{equation}
Once again, the matrix decouples into two blocks (rows 1,4,5 and rows
2,3,6). The first block is
\begin{equation}
  \check Q_{3A}=1+\frac{\sqrt T}{2}
  \left(\begin{array}[c]{ccc}
      \hat K_-(\chi) & -1 & 0\\
      -1 & 0 & e^{-i\chi\hat\tau_3} \\
      0 & e^{i\chi\hat\tau_3} & \hat K_+(0)
    \end{array}\right) \;.
\end{equation}
It is already evident, that we will encounter a three particle
process, if we apply the transformation $\check
U=$diag$(\exp(i\chi\hat\tau_3),\exp(i\chi\hat\tau_3),1)$. This yields
\begin{equation}
  \check U\check Q_{3A}\check U^\dagger =1+\frac{\sqrt{T}}{2}
  \left(\begin{array}[c]{ccc}
      \hat K_-(3\chi) & -1 & 0\\
      -1 & 0 & 1\\
      0 & 1 & \hat K_+(0)
    \end{array}\right) \;. 
\end{equation}
Evaluating the determinant we obtain the counting statistics (including
the other block, see below)
\begin{equation}
  S(\chi)=\frac{2eVt_0}{h}\ln\left(1+\frac{T^3}{(4-3T)^2}\left(e^{-i3\chi}-1\right)\right)\,.
\label{eq:fcs-thirdorder}
\end{equation}
Evidently this correspond to the binomial transfer of packages of three
charges, where the probability of a third order process is
$P_3=T^3/(4-3T)^2$. A similar procedure may be applied to the
second block $\check Q_{3B}$. The result is the same.  Physically, the
two blocks correspond to two independent processes which differ by the
spin. 

For higher order processes the calculation goes in complete analogy. The
property of a decoupling into two independent blocks remains. Further
more it is possible to the shift the entire $\chi$-dependence to the
uppermost (or the lowest) block. This is achieved by a series of unitary
operations of the type $(1,...,1,\exp(in\chi\hat\tau_3),1,...,1)$. One
can easily convince oneself, that for a process of order $N$ this gives
e.~g. the upper-left block $\hat K_-(N\chi)$ and the remaining matrix is now
independent of $\chi$. For example a 5th-order process yields \begin{equation}
  1+\frac{\sqrt{T}}{2}\left(
    \begin{array}[c]{ccccc}
      \hat K_-(5\chi) & 1 & 0 & 0 & 0 \\
      1 & 0 & 1& 0 & 0 \\
      0&  1 & 0 & 1 & 0\\
      0& 0 & 1 & 0 & 1 \\
      0 & 0 & 0 & 1 & \hat K_+(0)\\
    \end{array}\right)\,.
\end{equation}
Additionally, the signs of the off-diagonal element may be removed by
unitary transformations.  Evaluating the determinant we find
$S(\chi)= (2eVt_0/h)
\ln\left[1+P_5\left(e^{in\chi}-1\right)\right]$, where
$P_5=T^5/(16-20T+5T^2)^2$. This expression describes binomial transfers
of $5$ charges with probability $P_5$.

Using the above scheme, it is also possible to derive recursion
relations for the probabilities. We find the probability for a process
of order N
\begin{equation}
  P_N = \frac{1}{
    1+\frac{
      \left(\left(1+\frac{\sqrt{T}}{2}\right)\alpha_{N-1}^+-\frac{T}{4}\right)
      \left(\left(1-\frac{\sqrt{T}}{2}\right)\alpha_{N-1}^--\frac{T}{4}\right)}{
      \sqrt{T}\frac{T}{4} \gamma_{N-1}}}\,.
\end{equation}
The factors $\alpha_\pm$ and $\gamma$ are determined from the recursion
relations
\begin{equation}
  \alpha^{\pm}_n=1-\frac{T}{4\alpha^{\pm}_{n-1} }
  \quad,\quad
  \gamma_n=\frac{T}{4}\frac{\gamma_{n-1}}{\alpha^+_{n-1}\alpha^-_{n-1}}\,,
\end{equation}
with the initial conditions
\begin{equation}
  \gamma_1=\sqrt{T}\quad,\quad
  \alpha^\pm_1=1\pm\frac{\sqrt{T}}{2}\,.
\end{equation}
For general subharmonic voltages $2\Delta/(N-1)$ we find the counting
statistics
\begin{equation}
S(\chi)=\frac{2eVt_0}{h}
\ln\left[1+P_n\left(e^{in\chi}-1\right)\right]\,,
\end{equation}
where the probabilities are given by
\begin{eqnarray}
  \label{eq:mar-probabilities}
  P_2 & = & \frac{T^2}{(2-T)^2}\,, \\
  P_3 & = & \frac{T^3}{(4-3T)^2} \nonumber\,,  \\
  P_4 & = & \frac{T^4}{(8-8T+T^2)^2} \nonumber\,,  \\
  P_5 & = & \frac{T^5}{(16-20T+5T^2)^2} \nonumber\,,  \\
  P_6 & = & \frac{T^6}{(2-T)^2(16-16T+T^2)^2} \nonumber\,,  \\
  P_7 & = & \frac{T^7}{(64-112T+56T^2-7T^3)^2} \nonumber\,,  \\
  P_8 & = & {\frac {{T}^{8}}{ \left( {T}^{4}-32\,{T}^{3}+160\,{T}^{2}-256\,T+128
      \right) ^{2}}}\,.\nonumber
\end{eqnarray}
Note the limiting cases of these probabilities $P_N\sim T^N/4^{N-1}$
for $T\ll 1$ and $P_N=1$ for $T=1$. 

We can draw several conclusions from the toy model. First we obtain
simple expressions for the probabilities of multiple charge $P_N$, which
are not simple products of Andreev reflection probabilities and
quasiparticle transmissions, see
Eq.~(\ref{eq:mar-probabilities}). Furthermore it is interesting to note
that by virtue of the unitary transformations we can interpret the
charge transfer as simultaneous transmission of $N$ quasiparticles. This
explanation does not invoke any kind of combined transfer of
Cooper pairs and quasiparticle.

\subsection{Full expressions}

Let us now discuss the full expression of the probabilities $P_n(E,V)$ at
zero temperature. Since $\check Q$ has a block-tridiagonal form,
in order to calculate its determinant we can use the a recursion
technique similar to the one describe for the toy model.
We define the following 4$\times$4 matrices
\begin{eqnarray}
\check F_{\pm n} & = & \check Q_{\pm n,\pm n} - \check Q_{\pm n,\pm n\pm 2}
\check F^{-1}_{\pm n \pm 2} \check Q_{\pm n \pm 2,\pm n}  \;;\; n \ge 2
\nonumber \\
\check F_0 & = & \check Q_{0,0} - \check Q_{0,-2} \check F^{-1}_{-2}
\check Q_{-2,0} - \check Q_{0,2} \check F^{-1}_{2} \check Q_{2,0} ,
\end{eqnarray}

\noindent
With these definitions, $\det \check Q$ is simply given by $\det \check Q =
\prod^{\infty}_{j=-\infty} \det \check F_{2j}$. In practice, $\det
\check F_{n} =1$ if $|n| \gg \Delta/|eV|$. This reduces the problem
to the calculation of the determinants of $4 \times 4$ matrices.

In the zero-temperature limit one can work out this idea analytically, 
and after very lengthy but straightforward algebra, we
obtain the following expressions for $P^{\prime}_n(E,V)$

\begin{widetext}
\begin{eqnarray}
\label{eq:probabilities}
P^{\prime}_n(E,V) & = & \sum^{n-1}_{l=0} J_{-n+l} 
\left[ \prod^{l-1}_{k=-n+l+1} (T/4) |f^A_k|^2 \right] 
J_l \;;\;\; n \ge 1 \nonumber  \\
P^{\prime}_0(E,V) & = & K
\left[Z_{0}^R \left( 1 + \frac{\sqrt{T}}{2} (g^R_0 - g^A_{-1}) - \frac{T}{4}
(f^A_{-1})^2 B^A_{-2} \right) - \frac{T}{4} (f^R_0)^2 \right]
\Bigg[ R \leftrightarrow A \Bigg]
\end{eqnarray}

\noindent
Here, we have used again the shorthand $g^{A,R}_n(E) \equiv g^{A,R}(E+neV)$, and defined

\begin{equation}
Z_{\pm n}^{\alpha}=1 \pm
\frac{\sqrt{T}}{2} (g^\alpha_{\pm (n+1)} - g^\alpha_{\pm n}) - \frac{T}{4}
(f^\alpha_{\pm (n+1)})^2 B^\alpha_{\pm (n+2)} \;\;;\;\; n \ge 0 \label{eq:z}\,,
\end{equation}
where $\alpha=R,A$, $K = (\prod_{j=1}^\infty \det \check F_{-2j})
(\prod_{j=1}^\infty \det \check F_{2j})$
and the different functions can be expressed as follows $(n \ge 0)$

\begin{eqnarray}
  \label{eq:coefficients}
  \left(B^{\alpha}_{\pm n}\right)^{-1} & = & 1 \pm \frac{\sqrt T}{2}
  (g^\alpha_{\pm n} - g^\alpha_{\pm (n-1)}) - \frac{T}{4}
  (f^\alpha_{\pm n})^2 / Z^\alpha_{\pm n}\,,\\
  \det \check F_{\pm n} & = &  \prod_{\alpha=A,R} \left[ Z^\alpha_{\pm n}
    ( 1 \pm \frac{\sqrt T}{2} (g^\alpha_{\pm n} - g^\alpha_{\pm (n-1)})
    ) - \frac{T}{4} (f^\alpha_{\pm n})^2 \right] \,, \nonumber \\
  J_{\pm n} & = & (\prod_{j=1}^\infty \det \check F_{\pm (n+ 2j)}) \left\{   
    \frac{\sqrt{T}}{2} (g^A_{\pm n} - g^R_{\pm n}) \left[
      Z^R_{\pm n} Z^A_{\pm n} - \frac{T}{4} |f^A_{\pm n}|^2 \right]
    \mp \frac{T}{4} (f^A_{\pm n} - f^R_{\pm n}) \left[ f^R_{\pm n}
      Z^A_{\pm n} + f^A_{\pm n} Z^{R}_{\pm n} \right] \right\}\,.\nonumber
\end{eqnarray}
\end{widetext}

Notice that, since at zero temperature the charge only flows in
one direction, only the $P_n$ with $n \ge 0$ survive.
It is worth stressing that the full information about the
transport properties of superconducting point contacts is
encoded in these probabilities. Let us also remark
that $P_n(E,V)$ are positive numbers bounded between 0 and 1,
and fulfill the normalization condition $\sum_n P_n(E,V) =
1$. Thus, we see that for the finite bias dc transport, where the
superconducting phase does not play any role, there is no problem with
the typical interpretation of $P_n$ as probabilities~\cite{Belzig2001}. 
Moreover, although at a first glance the expressions of 
Eqs.~(\ref{eq:probabilities}-\ref{eq:coefficients}) look complicated, 
they can be easily computed and provide the most efficient way to calculate
the transport properties of these contacts. In practice, to
determine the functions $B^{A,R}_{n}$ and $\det \check F_{n}$, one
can use the boundary condition $B^{A,R}_{n} = \det \check F_{n} =1$
for $|n| \gg \Delta/|eV|$. 

In view of Eqs.~(\ref{eq:probabilities}-\ref{eq:coefficients}) the
probabilities $P_n$ can be interpreted in the following way. $P_n$
is the probability of a MAR of order $n$, where a quasiparticle in
an occupied state at energy $E$ is transmitted to an empty state at
energy $E+neV$. The typical structure of the expression for 
this probability consists of the product of three terms. First,
$J_0$ gives the probability to inject the incoming quasiparticle at
energy $E$. The term $\prod^{n-1}_{k=1} (T/4) |f^A_k|^2$
describes the cascade of $n-1$ Andreev reflections, in which an
electron is reflected as a hole and vice versa, gaining an energy
$eV$ in each reflection. Finally, $J_n$ gives the probability to
inject a quasiparticle in an empty state at energy $E+neV$. 
This interpretation is illustrated in Fig.~\ref{MAR}, where we show
the first four processes for BCS superconductors. The product of
the determinants in the expression of $J_n$ (see
Eq.~(\ref{eq:coefficients})) describes the possibility that a
quasiparticle makes an excursion to energies below
$E$ or above $E+neV$. In the tunnel regime this
possibility is very unlikely and at perfect transparency is forbidden.
For this reason the expressions of the MAR probabilities simplify a lot
in these two limits, as we discuss in the next paragraphs.
 
In the tunnel regime a perturbative calculation yields $(n \ge 1)$

\begin{equation}
P_n (T \ll 1) = \frac{T^n}{4^{n-1}} \rho_0 \rho_n \prod^{n-1}_{k=1}
|f^A_k|^2 \;,
\label{eq:tunnel}
\end{equation}

\noindent
where $\rho(E)$ is the reservoir density of states. If we use
this result in the current expression (see below), we recover exactly the result
of the multiparticle tunneling theory of Schrieffer and Wilkins~\cite{MPT}.
As we mentioned in the introduction, the expression above leads to divergences
in the current, which shows that this problem is non-perturbative in
the transmission. Thus, even at low transparencies one has to use
the full expression of Eqs.~(\ref{eq:probabilities}-\ref{eq:coefficients}),
where the mentioned divergences are regularized in a natural manner.

For perfect transparency $(T=1)$, the absence of normal backscattering
makes the expressions of the probabilities $P_n(E,V)$ much simpler, and
one can show that they can be written as $(n \ge 1)$

\begin{equation}
P_n(T=1) = \sum^{n-1}_{l=0} (1 - |a_{-n+l}|^2) \left[ \prod^{l-1}_{k=-n+l+1}
|a_k|^2 \right] (1 - |a_l|^2) ,
\label{eq:balistic}
\end{equation}

\noindent
where $a(E)$ is the Andreev reflection coefficient defined as
$a(E) = -if^R(E) / \left[ 1 + g^R(E) \right]$, and $a_n = a(E+neV)$.
As can be seen in Eq.~(\ref{eq:balistic}), a quasiparticle
can only move upwards in energy due to the absence of normal reflection.
If we use this expression in the current formula we recover the result obtain
by Klapwijk, Blonder and Tinkham~\cite{KBT} for $T=1$.

\section{Application to different situations}

As explained in the previous section, with the expression of the MAR probabilities
we can easily describe many different transport properties. Moreover, notice that
so far we have not made any assumption about the leads Green's functions $g^{A,R}$ 
and $f^{A,R}$ entering in the expressions of $P_n(E,V)$. Therefore, these 
expressions allow us to address a great variety of situations. In this section
we analyze the zero-temperature transport properties of three different situations: 
(i) a contact between BCS superconductors, (ii) a contact between superconductor
under the influence of pair-breaking mechanisms and (iii) a short diffusive
SNS contact, where N is a normal disordered region shorter than the superconducting
coherence length.

\subsection{BCS superconductors}

\begin{figure}[tb]
\includegraphics[width=0.9\columnwidth,clip=]{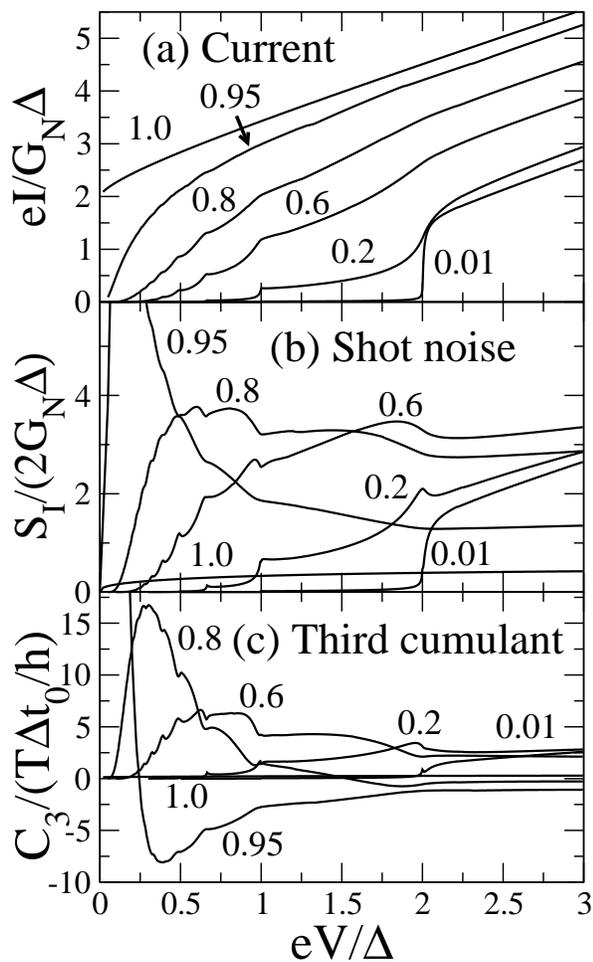}
\caption{\label{bcs-currents} Current, shot noise and third cumulant at
zero temperature as a function of the voltage for BCS superconductors
of gap $\Delta$. The different curves correspond
to different transmission coefficients as indicated in the panels.
Here, $G_N = (2e^2/h)T$ is the normal state conductance.
}
\end{figure}

\begin{figure}[tb]
\includegraphics[width=0.9\columnwidth,clip=]{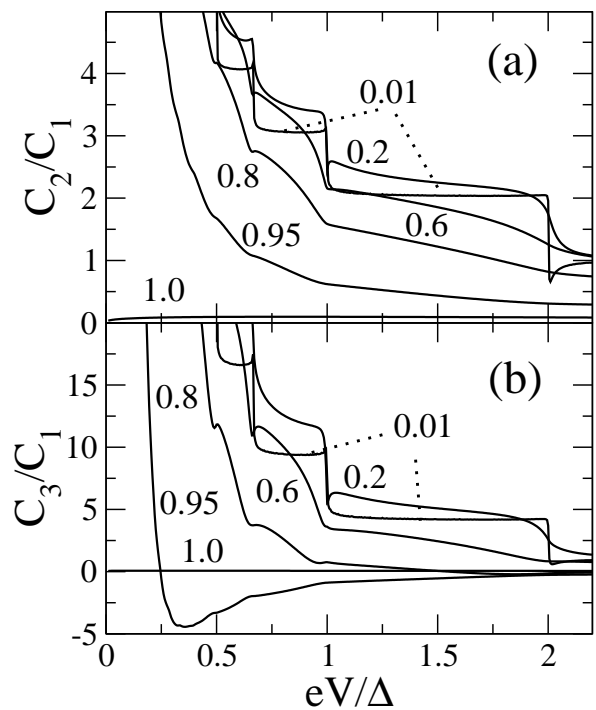}
\caption{\label{bcs-cumulants} (a) Second cumulant and (b) third cumulant
at zero temperature for BCS superconductors. Both are normalized to the
first cumulant (the average current). The transmissions are indicated in the plots.}
\end{figure}

Let us start analyzing the most standard situation, namely a contact between
two BCS superconductors with a gap $\Delta$. In this case $f^{A,R} = i\Delta
/ \sqrt{(E \mp i \eta)^2 - \Delta^2}$, where $\eta = 0^+$, and $g^{A,R}$
follows from normalization. As mentioned in the introduction the current and 
noise of such a contact have been thoroughly studied both 
theoretically~\cite{Bratus1995,Averin1995,Hurd1996,Cuevas1996,Cuevas1999,Naveh1999} 
and experimentally~\cite{Scheer1997,Scheer1998,Ludoph2000,Goffman2000,Cron2001}. 
Our goal here is to show how the knowledge of the FCS provides a
new and deeper insight into the different transport properties.

In Fig.~\ref{bcs-currents} we show the first three cumulants of the
charge transfer distribution: current, shot noise and skewness (third cumulant).
Let us discuss their most remarkable features. (i) The current exhibits
the so-called subharmonic gap structure, as discussed in the introduction.
This subgap structure evolves from a step-like behavior for low transmission to
its disappearance at perfect transparency. (ii) The shot noise in the subgap
region can be much larger than the Poisson noise ($S_{I,Poisson}=2eI$). Moreover,
in the tunneling regime the effective charge defined as the ratio $q \equiv S_I/2I$ is
quantized in units of the electron charge: $q(V)/e = 1 + \mbox{Int}(2\Delta/eV)$. 
This is illustrated in Fig.~\ref{bcs-cumulants}, where the ratios $C_2/C_1$ and 
$C_3/C_1$ are shown as a function of the voltage. (iii) As shown in 
Fig.~\ref{bcs-cumulants}, the third cumulant at low transmissions is described 
by $C_3 = q^2 C_1$, where again $q$ is the quantized effective charge defined above. 
For higher transmissions this cumulant is negative at high voltage
as in the normal state, where $C_3 = (t_0/h) T(1-T)(1-2T) eV$, but it becomes
positive at low bias, and after this sign change there is a huge increase of 
the ratio $C_3/C_1$.

\begin{figure}[tb]
\includegraphics[width=0.9\columnwidth,clip=]{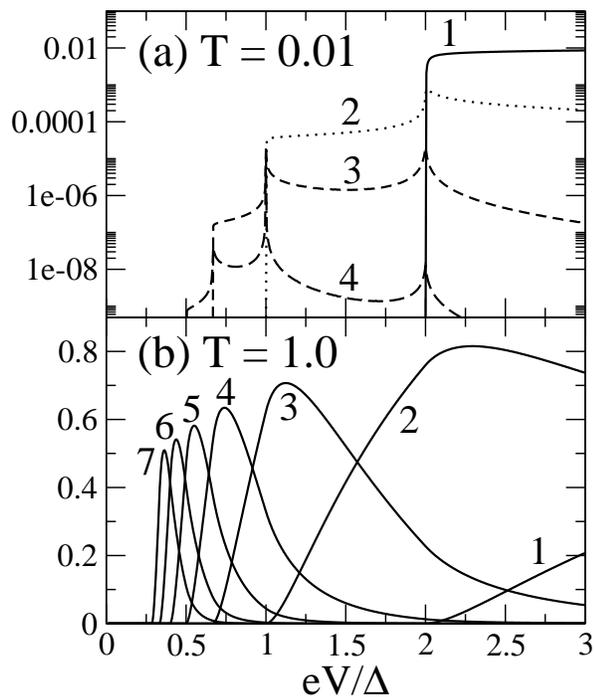}
\caption{\label{bcs-probabilities} Average MAR probabilities 
$\bar P_n(V) \equiv (1/eV) \int^{eV}_0 dE \; P_n(E,V)$ as a function
of voltage for a contact between BCS superconductors at zero temperature.
The two panels correspond to two different transmissions. The index of
the processes is indicated in the plots. Notice the logarithmic scale
in the panel (a). }
\end{figure}

The features described in the previous paragraph can be easily
understood with the help of an analysis of the probabilities $P_n(E,V)$.
To give an idea about them, in Fig.~\ref{bcs-probabilities} we have
plotted their average value defined as $\bar P_n(V) \equiv (1/eV)
\int^{eV}_0 dE \; P_n(E,V)$ for two very different transmissions. First
of all, notice that, no matter what the transmission is, the probability
of an n-order MAR has a threshold voltage $eV_n = 2\Delta/n$, below
which the process is forbidden. When $V> V_n$ an n-order MAR gives a new
contribution to the transport, which is finally the explanation of the
subharmonic gap structure.  On the other hand, the big difference
between the tunneling regime and perfect transparency can be explained
as follows. At low transparency there are two factors that make the
subgap structure so pronounced. First, at $V_n$ the n-order MAR is a
process that connects the two gap edges, where the BCS density of states
diverges (see Eq.~(\ref{eq:tunnel})). This fact, together of course with
its higher probability, implies that this MAR rapidily dominates the
shape of the I-V curves giving rise to a non-linearity at $V_n$. Second,
at $V_n$ there is a huge enhancement of the probabilities of the MARs of
order $m > n$.  This is due to the fact that precisely at $V_n$ the MAR
trajectories can connect both gap edges, which as can be seen in
Eq.~(\ref{eq:tunnel}) increases enormously their probability. At perfect
transparency, the MAR probabilities do not exhibit any abrupt feature
(see Fig.~\ref{bcs-probabilities}b). This is due to the fact that the
BCS density of states is renormalized, and in particular, the
divergences disappear (see Eq.~(\ref{eq:balistic})). This fact explains
naturally why the subharmonic gap structure is completely washed out at
$T=1$.

Another interesting feature of the MAR probabilities occurs at low transparencies.
As one can see in Fig.~\ref{bcs-probabilities}a, at a voltage $2\Delta/n < eV < 
2\Delta/(n-1)$ the MAR of order $n$ has a much higher probability than the other
MARs. This means that in this voltage window the n-order MAR clearly dominates 
the transport properties and the charge is predominantly transferred in packets 
of $ne$. This fact explains the charge quantization in the tunnel regime observed
both in $C_2$ and $C_3$ (see Fig.~\ref{bcs-cumulants}). More generally, this
fact implies that at low transparencies the multinomial distribution of Eq.(\ref{eq:marfcs})
becomes Poissonian, and in this limit all the cumulants are proportional to
the current: $C_n = (q(V)/e)^n C_1$, where $q(V)$ is the voltage-dependent quantized
charge.  When the transmission is not very low, there are always several MARs 
that give a significant contribution to the transport at every voltage (see 
Fig.~\ref{bcs-probabilities}b). This explains why the charge is in general
not quantized.

\begin{figure}[t]
\includegraphics[width=0.8\columnwidth,clip=]{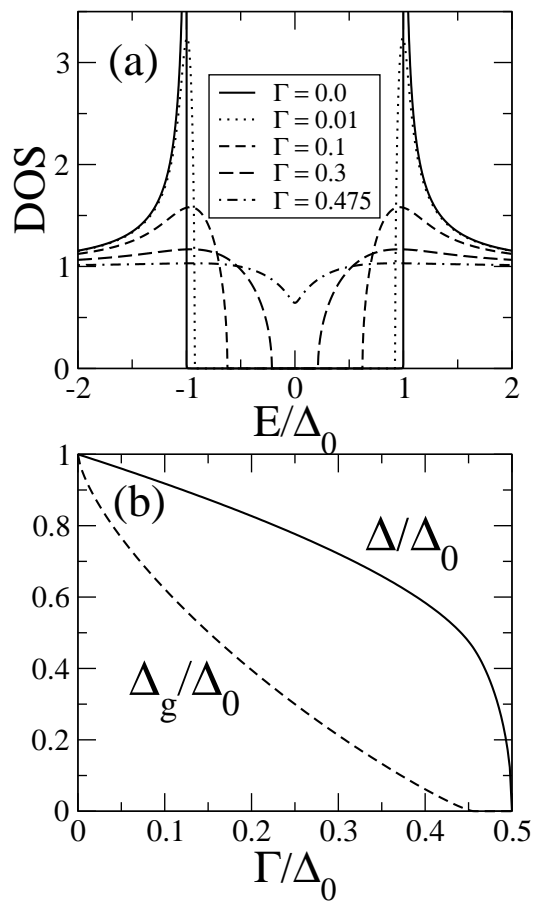}
\caption{\label{pb-dos} (a) Density of states as a function of energy of a 
superconductor for different values of the depairing energy $\Gamma$
measured in units of gap in the absence of pair-breaking $\Delta_0$.
(b) Order parameter $\Delta$ and spectral gap $\Delta_g$ in units of 
$\Delta_0$ as a function of the depairing energy $\Gamma$ normalized by
$\Delta_0$.}
\end{figure}

The explanation for the sign change of $C_3$ at low bias and high transparencies
can be found in Eq.~(23). In order to get a positive value for 
$C_3$, one needs the first two terms in Eq.~(23) to dominate, 
which happens when $P_n \ll 1$. This is precisely what happens at low bias, where
the MAR probabilities are rather small. On the other hand, the huge enhancement 
after the sign change is due to fact that $n$, the charge transferred by these 
MARs, is indeed huge at low bias.

Finally, at $T=1$ the cumulants $C_n$ (with $n > 1$) do not completely vanish
due to the fact that at a given voltage different MARs give a significant contribution,
and therefore their probability is smaller than one (see Fig.~\ref{bcs-probabilities}(b)).

\subsection{Pair-breaking mechanisms}

It is well-known that there are many mechanisms that can lead to pair-breaking
effects, which modify the quasiparticle spectrum of a superconductor. Typical
examples are a magnetic field, supercurrents or magnetic impurities. It was shown
in the 1960's that for diffusive superconductors various pair-breaking mechanisms
can be described in a unified manner with a single parameter $\Gamma$, the depairing 
energy, which describes the strength of the pair-breaking~\cite{Maki1969}. The only 
difference between these mechanisms is contained in the microscopic expression of $\Gamma$.
For instance, for thin a film of thickness $d$ much smaller than the superconducting
coherence length in a magnetic field $H$ parallel to the film $\Gamma = 
D e^2 d^2 H^2/(6\hbar c^2 )$, where $D$ is the diffusion constant. In these situations
the energy-dependent retarded Green's function can be calculated from~\cite{Maki1969}

\begin{figure}[tb]
\includegraphics[width=\columnwidth,clip=]{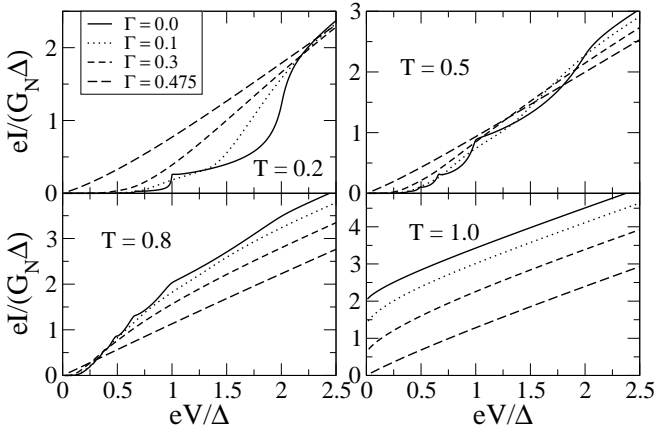}
\caption{\label{pb-current} Zero temperature current-voltage characteristics
 for superconductors with a depairing energy $\Gamma$ in units of
 $\Delta_0$. The current and the voltage have been normalized with the
 order parameter $\Delta$ at the corresponding $\Gamma$. The different panels
 correspond to different transmissions values.}
\end{figure}

\begin{equation}
g^R  =  \frac{u}{\sqrt{u^2 -1}} = u f^R \;,\; \frac{E}{\Delta} =
u \left[ 1 - \frac{\Gamma}{\Delta \sqrt{1-u^2}} \right] .
\end{equation}

\noindent
Here, $\Delta$ is the order parameter, which is in this case differs from the 
spectral gap and it has to be determined self-consistently~\cite{Skalski1964}. For
small $\Gamma$ the pair-breaking mechanisms result in a smearing of the BCS 
singularities in the density of states and in a suppression of the spectral
energy gap $\Delta_g$ to a reduced value
$\Delta_g = \Delta \left[ 1 - (\Gamma/\Delta)^{2/3} \right]^{3/2}$. The gap disappears
completely at $\Gamma \approx 0.45\Delta_0$, where $\Delta_0$ is the order parameter in
the absence of pair-breaking. The gapless superconductivity survives until the
critical value $\Gamma_C = 0.5\Delta_0$. This behavior is illustrated in 
Fig.~\ref{pb-dos}(a), where we show the density of states as a function of energy
for different values of $\Gamma$ in units of $\Delta_0$. In Fig.~\ref{pb-dos}(b)
one can see the evolution of the order parameter and spectral gap with the 
depairing energy.

\begin{figure}[tb]
\includegraphics[width=\columnwidth,clip=]{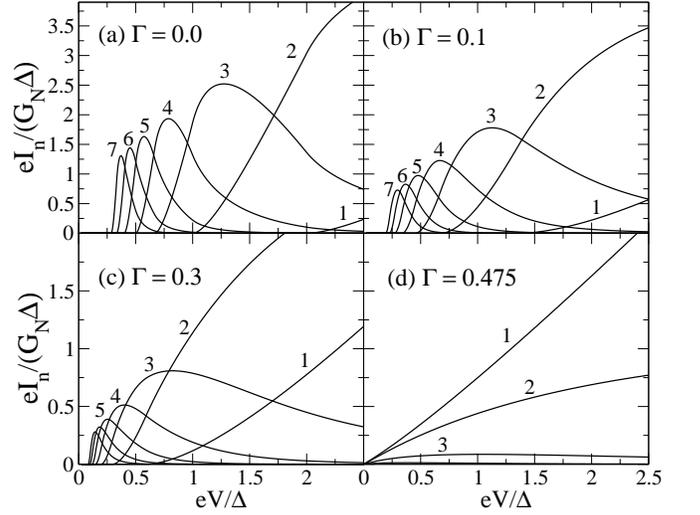}
\caption{\label{pb-T=1.0} Current contribution of processes $n=1,2,...$ for
$T=1$ as a function of voltage for superconductors with a depairing energy $\Gamma$ 
in units of $\Delta_0$. The current and the voltage have been normalized with the
order parameter $\Delta$ at the corresponding $\Gamma$. The order of the processes
is indicated in the plots.}
\end{figure}

\begin{figure}[tb]
\includegraphics[width=\columnwidth,clip=]{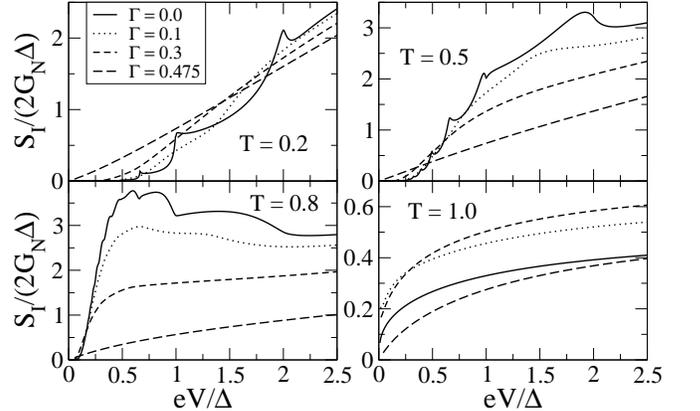}
\caption{\label{pb-noise} Zero temperature noise for superconductors with
a depairing energy $\Gamma$ in units of $\Delta_0$. The current and the voltage 
have been normalized with the order parameter $\Delta$ at the corresponding $\Gamma$.
The different panels correspond to different transmissions values. }
\end{figure}

\begin{figure}[tb]
\includegraphics[width=\columnwidth,clip=]{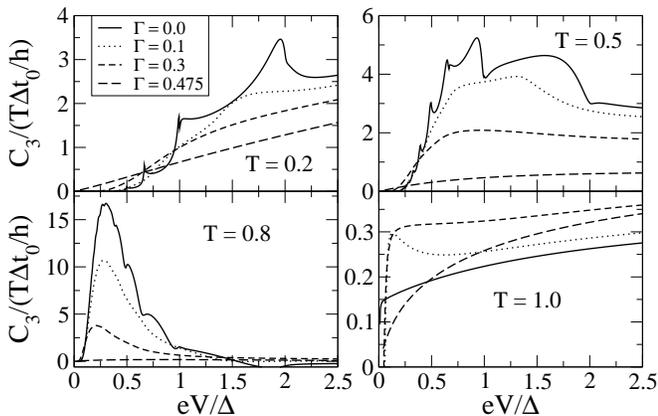}
\caption{\label{pb-c3} Zero temperature third cumulant for superconductors with
a depairing energy $\Gamma$ in units of $\Delta_0$. The current and the voltage
have been normalized with the order parameter $\Delta$ at the corresponding $\Gamma$.
The different panels correspond to different transmissions values.
}
\end{figure}

Let us discuss now how this modified density of states is reflected in
the transport properties. In Fig.~\ref{pb-current} we show I-Vs for
different transmissions and different values of the depairing energy.
The most noticeable features are: (i) the subharmonic gap structure is
shifted to voltages $eV = 2\Delta_g/n$, and (ii) the subgap structure
progressively disappears as the pair-breaking strength is increased.
These features are simple consequences of the evolution of the density
of states with $\Gamma$. Anyway, one can get a further insight by
analyzing the contribution to the current of the individual MAR
processes: $I_n = (2e/h) \int^{eV}_0 dE\; P_n(E,V)$.  These quantities
are plotted in Fig.~\ref{pb-T=1.0} for $T=1$. As one can see, the
threshold voltage for a n-order MAR is now $eV_n = \Delta_g/2n$ as a
consequence of the reduced spectral gap. As the gap diminishes, the
processes of lowest order dominate the I-Vs even at low bias. It
is interesting to notice that even in a gapless situation $(\Gamma =
0.475)$ there is a finite contribution of the MARs. It is worth
mentioning that in Refs.~\cite{Scheer2000} and \cite{Bascones2000} the
type of theory described here accounted for the magnetic field
dependence of the I-Vs of atomic contacts.

Let us turn now our attention to the second and third cumulants that can
be seen in Fig.~\ref{pb-noise} and Fig.~\ref{pb-c3}, respectively. As in
the case of the current, the subharmonic gap structure is shifted and
smoothed as the gap evolves with $\Gamma$. Moreover, one can notice that
for high transparencies and in the subgap region there is a great reduction
of both cumulants as $\Gamma$ increases. This is a consequence of the
fact that low order MARs dominate even at low bias, which in practice means
that the charge transferred at these voltages is on average not very big.

\subsection{Diffusive SNS contacts}

So far we have discussed the case of a single channel contact. The results
are trivially generalized to the multichannel case by introducing a sum
over the conduction channels. In this subsection we briefly address the case of a 
short diffusive SNS junction with a large number of transmission channels
and diffusive electron transport in the normal N region. The superconducting
leads are considered as BCS superconductors. In this case,
the distribution of transmission coefficients is continuous, and it is
characterized by the density function $\rho(T)$, which has the well-known
bimodal form~\cite{Nazarov1994}

\begin{equation}
\rho(T) = \frac{G_N}{2G_0} \frac{1}{T \sqrt{1-T}} \;,
\end{equation}

\noindent
where $G_N$ is the normal-state conductance of the N region ad
$G_0=2e^2/h$ is the conductance quantum. Then, the different cumulants
can be calculated from the single-channel results $C_n(T)$ as follows

\begin{equation}
C_n = \int^1_0 dT\; \rho(T) C_n(T).
\end{equation}

\begin{figure}[tb]
\includegraphics[width=\columnwidth,clip=]{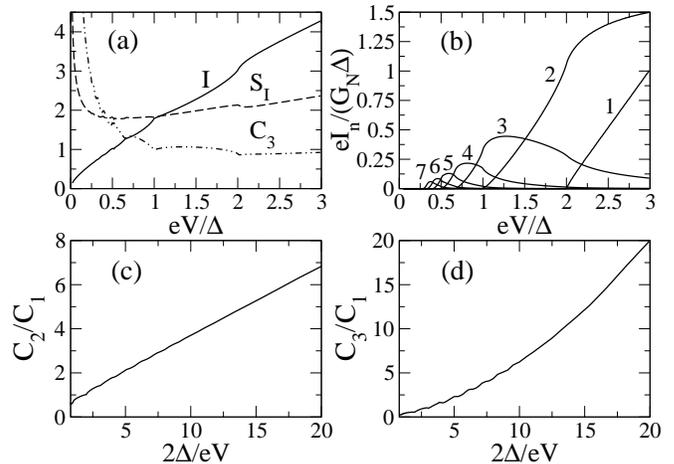}
\caption{\label{diffusive} Zero temperature transport properties
  of a short diffusive SNS junction. (a) First three cumulants: current
  in units of $(G_N \Delta/e)$, shot noise in units of $(2G_N\Delta)$
  and the third cumulant in units of $(G_N \Delta t_0/h G_0)$. (b)
  Current contribution of the different processes. (c) Ratio $C_2/C_1$
  as a function of the inverse of the voltage. (d) Ratio $C_3/C_1$ as as
  a function of the inverse of the voltage.}
\end{figure}

In Fig.~\ref{diffusive}(a) we show the first three cumulants for this 
SNS system. Both the current and the noise have previously discussed 
in the literature~\cite{Bardas1997,Naveh1999}, and here we recover these
results. Both quantities exhibit a subharmonic gap structure which is
a result of the competition of channels with different transparencies.
Again, this structure can be understood by analyzing the individual
contributions to the current of the different MARs, see Fig.~\ref{diffusive}(b).
As one can see, at every voltage there are several processes giving a 
significant contributions, which makes that subgap structure much smoother
than in the single-channel case. This fact also explains the absence of
the charge quantization in this multichannel case. This is illustrated
in Fig.~\ref{diffusive}(c), where we show the ratio $C_2/C_1$ as a measure
of the effective charge. Notice that at low bias this effective charge
grows as $(1/V)$ as obtained in Ref.~[\onlinecite{Naveh1999}]. In this
regime the numerical results can be approximately described by the following
linear function: $C_2/C_1 = 0.31 (2\Delta/eV) + 0.55$. On the other hand,
the third cumulant exhibits a huge increase at low voltages~\cite{Johansson2003}.
In particular, as shown in Fig.~\ref{diffusive}(d), the ratio $C_3/C_1$ grows
as $(1/V)^2$ at low bias. In this regime the ratio can be approximated by
$C_3/C_1 = 0.05 (2\Delta/eV)^2 + 0.5$.

\section{Transport properties at finite temperatures}

So far we have discussed the transport properties of superconducting
point contacts at zero temperature. In this section we shall investigate
the role of the temperature, which we shall denote as $T_e$. We focus our 
attention to the case of a single channel contact between BCS superconductors. At finite
temperature it is not easy to determine analytically the probabilities $P_n(E,V)$,
and in this case we have calculated them numerically. The idea goes as follows.
According to Eq.~(18) we need to calculate the coefficients $P^{\prime}_n(E,V)$,
which are simply the Fourier coefficients of the series in Eq.~(18), i.e. 
\begin{equation}
P^{\prime}_n(E,V) = \frac{1}{2\pi} \int^{2\pi}_0 d\chi \;  e^{-in \chi}
\det \check Q(\chi) \;.
\end{equation}

\noindent
Finally, $\det \check Q(\chi)$ is calculated numerically. Of course, if
one is only interested in the different cumulant, one can easily calculate
them by taking the numerical derivative of the CGF, see Eq.~\ref{eq:cumulants1}.

\begin{figure}[tb]
\includegraphics[width=\columnwidth,clip=]{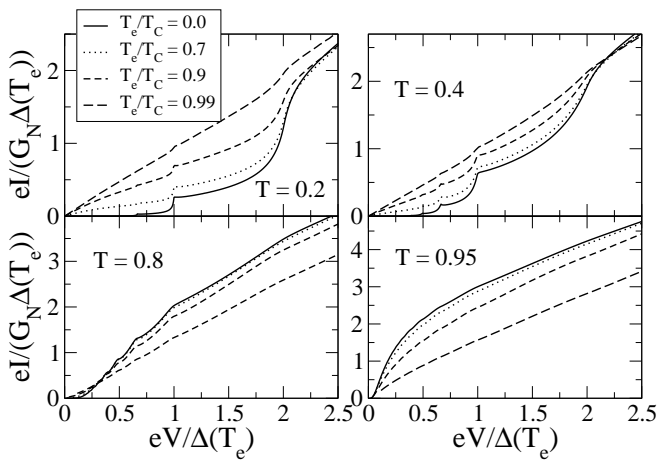}
\caption{\label{temp-current} Current-voltage characteristics at finite 
  temperature for BCS superconductors. The temperature is in units of
  the critical temperature $T_C$. The current and the voltage are
  normalized with the temperature-dependent gap.  The different panels
  correspond to different transmission values.}
\end{figure}

\begin{figure}[tb]
\includegraphics[width=\columnwidth,clip=]{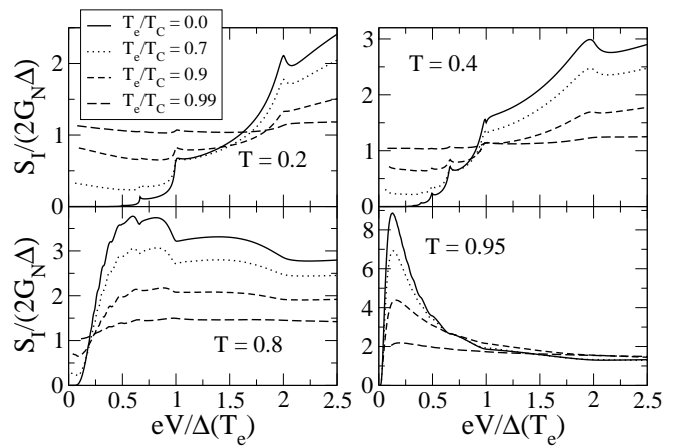}
\caption{\label{temp-noise} Finite temperature noise for BCS superconductors.
  The temperature is normalized with the critical temperature $T_C$. The
  different panels correspond to different transmission values. The
  voltage is normalized with the temperature-dependent gap, and the current
  with the zero-temperature gap. Note that the scaling is different from 
  the other plots in this section.}
\end{figure}

\begin{figure}[tb]
\includegraphics[width=\columnwidth,clip=]{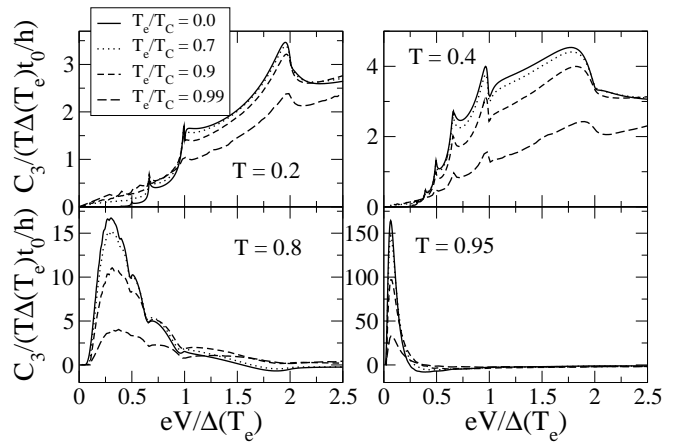}
\caption{\label{temp-c3} Finite temperature thrid cumulant for BCS superconductors.
  The temperature is normalized with the critical temperature $T_C$. The
  different panels correspond to different transmission values. The
  third cumulant and the voltage are normalized with the
  temperature-dependent gap.}
\end{figure}

\begin{figure}[tb]
\includegraphics[width=\columnwidth,clip=]{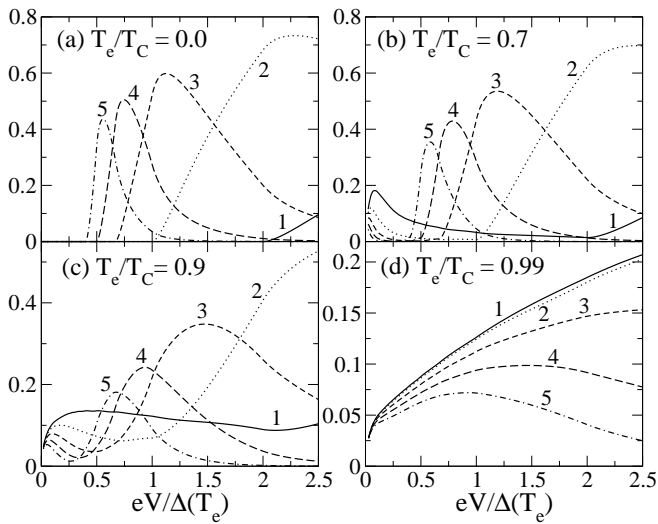}
\caption{\label{temp-probabilities} Average MAR probabilities 
  $\bar P_n(V) \equiv (1/eV) \int^{eV}_0 dE \; P_n(E,V)$ at finite
  temperature as a function of voltage for a contact between BCS
  superconductors with transmission $T=0.95$. The four panels correspond
  to different temperatures $T_e$ expressed in units of the critical
  temperature $T_C$. The index of the processes is indicated in the
  plots. }
\end{figure}

In Figs.~\ref{temp-current},\ref{temp-noise}, and \ref{temp-c3} we show
the current, noise and third cumulant, respectively, for different
transmission and temperatures ranging from zero to the critical one.
Notice that in order to get rid of the trivial temperature dependence
due to the decrease of the gap we have normalized the voltage by the
temperature-dependent gap $\Delta(T_e)$. As it can be seen in
Fig.~\ref{temp-current}, the temperature progressively smoothes the SGS
and increases the current for low transmissions. These are simple
consequences of the thermal excitation of quasiparticles.  For higher
transmissions the temperature has the opposite effect (see the lower two
panels in Fig.~\ref{temp-current}). The current decreases with
increasing temperature and approaches the normal state current-voltage
characteristic from above. At the same time the excess current, i.~e.
$I(V\gg\Delta/e)-G_NV$, vanishes obviously. So in short, by increasing
the temperature high-order Andreev reflections contribute less to the
current, which is dominated by thermally activated direct quasiparticle
tunneling. This behavior is clearly illustrated in
Fig.~\ref{temp-probabilities}, where we show the evolution with the
temperature of the average probability $\bar P_n(V) \equiv (1/eV)
\int^{eV}_0 dE \; P_n(E,V)$ of different processes for a contact with
transmission $T=0.95$. Notice that we only show the first electron
processes that give a positive contribution to the current.  Remember
that at finite temperature there are also hole processes that give a
negative contribution to the current, the magnitude of which is still
much smaller than the one of the electron processes in the shot noise
limit $eV\gg k_BT$. At vanishing voltages, of course, $P_n=P_{-n}$ as
required by the fluctuation-dissipation theorem.  

In Fig.~\ref{temp-probabilities} one can observe the following important
features. First, at finite temperature the different processes do not
have any finite threshold voltage, and they can contribute down to zero
bias due to thermal activation.  Second, as the temperature increases
the probability of the single quasiparticle processes is greatly
enhanced inside the gap. This fact results in a reduction of the average
effective charge transmitted through the contact. Finally, notice that
although the MAR probabilities are reduced inside the gap at finite
temperature, high-order processes can give a significant contribution to
the transport even at voltages larger than the gap at the corresponding
temperature. This is clearly at variance with the zero temperature case.
To understand this behavior, let us recall that the total voltage gain
for an order $n$ process is $neV$, which means essentially that higher
order processes can start well below the gap and end well above the gap.
Now, at finite temperature e.g. the end states above the gap are filled
with finite probability $f(E+neV)$, assuming that the process has
started with a quasiparticle at energy $E$. A certain process can only
happen if its final state is empty. This gives a factor $1-f(E+neV)$,
which enhances the chance for higher order processes, since they have to
end up at higher energies, for which this factor is larger . On the
other hand, a similar argument can be made about the initial state,
which has to be filled for the process to take place. Again, this is
more likely for higher order processes, since they can emerge from
energies well below the gap, which are completely filled also
at finite temperature.

It is interesting to discuss the qualitative different temperature
behavior of the second and third cumulants. The noise exhibits a
transition from pure shot noise at zero temperature to thermal noise
when the temperature is larger than the voltage. As it can be seen in
Fig.~\ref{temp-noise}, this transition is reflected in a saturation of
the noise at low bias to a finite value, which is given by the
fluctuation-dissipation theorem. It is interesting to note, that the
noise \textit{decreases} as a function of voltage in the transition
region from thermal to shot noise also for relatively small
transmissions. Such a behavior can be attributed to the multinomial
distribution. Interestingly, from Eq.~(\ref{eq:c2nm}) we see that the
correlations between processes of orders with opposite sign (e.~g
$m=-n$) tend to increase the noise. As these terms appear only if the
respective probabilities are non-negligible, the reduction of noise
below the thermal level can be understood as consequence of the
vanishing cross correlations between processes of orders with different
signs.

\begin{figure}[tb]
\includegraphics[width=0.9\columnwidth,clip=]{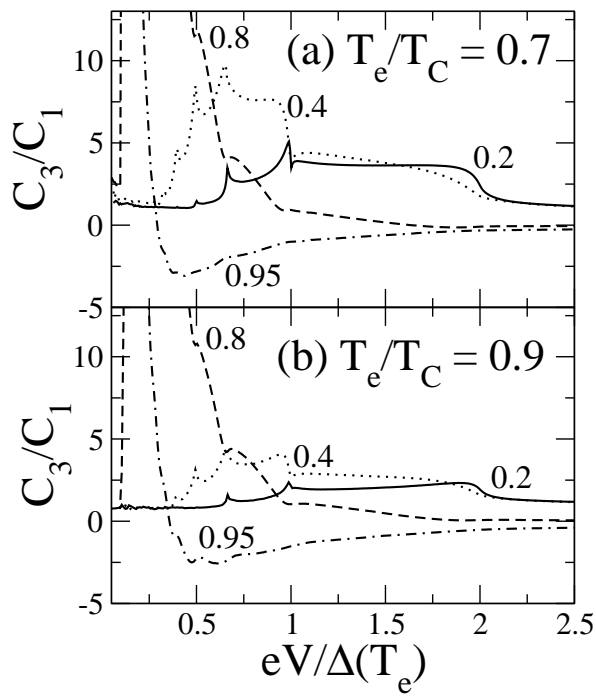}
\caption{\label{temp-ratio} Ratio $C_3/C_1$ for two different temperatures
as a function of the voltage for a contact between BCS superconductors.
The different curves correspond to different transmissions as indicated 
in the plots.}
\end{figure}

The temperature dependence of the third cumulant is very interesting.
First we recall that the third cumulant vanishes at zero voltage for any
temperature (as all odd cumulants do). In Ref.~\cite{Levitov2001} the
temperature dependence of the third cumulant for a quantum contact
between normal metals was calculated. It was shown, that an increasing 
transparency has quite a dramatic effect on the third cumulant. For a tunnel
junction (i.~e. for small transmission) $C_3$ is independent of the temperature
and it is simply equal to the $q^2 C_1$. However, this is interesting
because it allows a direct measurement of the charge $q$ transfered in an
elementary event even for voltages below the shot noise limit. Note,
that this relation holds also for non-linear current-voltage
characteristics, since it is a consequence of the bidirectional Poisson
distribution in this limit. The effects of a finite transparency are
even more dramatic. The third cumulant has a marked temperature
dependence, crossing over from a $FI$ dependence, where $F=1-T$ is the
Fano factor, to a novel high-temperature dependence $\sim FI(1-2T)$,
which can even become negative for $T>1/2$. In view of these findings,
we will now discuss our results for the temperature dependence of the
third cumulant of a superconducting point contact.

First, we note that in Fig.~\ref{temp-c3} $C_3$ has a temperature 
dependence even in the tunnel regime. As explained in the previous paragraph,
this in contrast with the normal state, where $C_3$ is almost independent
of the temperature, as it has been discussed theoretically in 
Ref.~\cite{Levitov2001} and observed experimentally in Ref.~\cite{Reulet2003}.
In our case the temperature dependence is due to the change in the MAR
probabilities caused by the thermal activation. As explained above, the 
thermal activation enhances the probability of the tunneling of single
quasiparticles inside the gap, which in turn reduces the average effective
charge. A consequence of this fact is the great reduction of the ratio
$C_3/C_1$ as the temperature increases. This is illustrated in 
Fig~\ref{temp-ratio}. This reduction is specially dramatic in the
subgap region for high transparencies, as it can be seen directly in 
Fig.~\ref{temp-c3}.

\section{Conclusions}

We have presented a detailed analysis of the full counting statistics in
superconducting point contacts at finite bias voltage. We have demonstrated
that the charge transfer in these systems is described by a multinomial
distribution of processes, in which multiple charges $ne$ (with $n=1,2,3,
...,20, ...$) are transferred through
the contact. These processes are nothing but multiple Andreev reflections.
The knowlegde of the full counting statistics allows us to obtain the 
probabilities of the individual MARs, providing so a deep insight into 
the electronic transport of these junctions. From the knowledge
of these probabilities one can easily calculate not only the current or the 
noise, but all the cumulants of the current distribution. We have also
shown that one can obtain analytical expressions for the MAR probabilities
at zero temperature, which provides the most efficient method to calculate 
the transport properties of these contacts. Moreover, the FCS approach 
allows us to describe a great variety of situations in a unified manner.

In this sense, we have addressed different situations such as contacts
between BCS superconductors, junctions between superconductors where a
pair-breaking mechanism ist acting or short diffusive SNS contacts.  We
have also discussed the temperature dependence of the first cumulants
and illustrated their peculiarities as compared with the normal case.
It is also worth mentioning that the formalism developed in this work
can be easily applied to other situations not addressed here such as
point contacts with proximity-effect superconductors~\cite{Scheer2001}
and Josephson junctions of unconventional
superconductors~\cite{Poenicke2002,Cuevas2002b}. 

From the full counting statistics view, we have found a new distribution
occuring in superconducting point contacts. The statistics takes the
form of a multinomial distribution of charge transfers of all orders,
which are allowed by the applied bias voltage. We have shown, that the
limit of opaque contacts provides an interesting situation, in which
Poissonian statistics makes it possible to observe multiple charge
transfers in a direct manner. Furthermore, we have discussed
consequences of the multinomial statistics of charge transfers of
different sizes at the same time. For example, an open contact has a
finite noise due to the presence of different MAR processes at the same
time. The temperature dependence of the counting statistics provides a
new insight in the transport characteristic, since we have shown that
higher order Andreev processes contribute also at voltages much larger
than the superconducting gap.

Finally we remark, that the FCS approach provides a fresh view of the
electronic transport of superconducting point contacts and it is seems
to be a natural choice for future theoretical analyses. On the other
side, superconducting contacts show an interesting new counting
statistics, namely a multinomial distribution, and we expect further
intersting results in other superconducting systems out of equilibrium.

We acknowledge discussions with A.~Levy Yeyati, A. Mart\'{\i}n-Rodero and
Yu.V.~Nazarov. JCC was financially supported by the DFG within the CFN and 
by the Helmholtz Gemeinschaft within the Nachwuchsgruppe program (contract 
VH-NG-029), and WB by the Swiss NSF and the NCCR Nanoscience.

{}
\end{document}